# A novel layered topology of auxetic materials based on the tetrachiral honeycomb microstructure

Ferdinando Auricchio[1], Andrea Bacigalupo[2]*, Luigi Gambarotta[3], Marco Lepidi[3], Simone Morganti[1], Francesca Vadalà[3]

[1]Department of Civil Engineering and Architecture, University of Pavia, Italy
[2]IMT School for Advanced Studies Lucca, Italy
[3]Department of Civil, Chemical and Environmental Engineering, University of Genova, Italy

**Abstract**

Microstructured honeycomb materials may exhibit exotic, extreme and tailorable mechanical properties, suited for innovative technological applications in a variety of modern engineering fields. The paper is focused on analysing the directional auxeticity of tetrachiral materials, through analytical, numerical and experimental methods. Theoretical predictions about the global elastic properties have been successfully validated by performing tensile laboratory tests on tetrachiral samples, realized with high precision 3D printing technologies. Inspired by the kinematic behaviour of the tetrachiral material, a newly-design bi-layered topology, referred to as bi-tetrachiral material, has been theoretically conceived and mechanically modelled. The novel topology virtuously exploits the mutual collaboration between two tetrachiral layers with opposite chiralities. The bi-tetrachiral material has been verified to outperform the tetrachiral material in terms of global Young modulus and, as major achievement, to exhibit a remarkable auxetic behaviour. Specifically, experimental results, confirmed by parametric analytical and computational analyses, have highlighted the effective possibility to attain strongly negative Poisson ratios, identified as a peculiar global elastic property of the novel bi-layered topology.

**Keywords:** Elastic properties, Experimental tests, Additive manufacturing, Finite element analysis, Beam lattice, Bi-tetrachiral material

---

* Corresponding Author

**INTRODUCTION**

Architectured composite materials characterized by a cellular honeycomb microstructure are experiencing an increasing success in a variety of innovative technological applications, by virtue of superior and tailorable physical-mechanical properties [1],[2],[3]. Among the other cellular topologies, honeycomb materials based on chiral and anti-chiral microstructure can offer peculiar extreme features of directional auxeticity, frequently conjugated with functional static performances of high resistance to fracture and indentation, atypical bending with synclastic curvatures, enhanced strength against buckling instabilities [4],[5],[6],[7],[8],[9]. Furthermore, the fine parametric tunability of the microstructural inertia and stiffness allows the systematic employment of chiral and anti-chiral materials to realize efficient and versatile phononic filters, elastic waveguides and acoustic diodes [10],[11],[12],[13],[14]. Within this challenging framework, the optimal design of the micromechanical properties opens interesting research perspectives towards the theoretical conceptualization and experimental validation of a new generation of smart materials targeted at innovative engineering applications, including impact absorption, negative refraction, shape morphing, wave trapping, vibration shielding, noise silencing and invisibility cloaking [15],[16],[17],[18],[19],[20].

Focusing the attention on the modern scientific literature about quasi-static laboratory tests of architectured materials, a pioneering investigation has experimentally verified the theoretical possibility to achieve strong auxetic properties (quantified by negative Poisson ratios close to unity) in medium-size samples of planar hexachiral honeycombs hand-built in polystyrene [21],[22]. Chiral and anti-chiral topologies of planar honeycombs have been analyzed by means of quasi-static experimental tests on small-size samples manufactured using selective laser sintering rapid prototyping of nylon powder. The experimental results have been successfully compared with the static response simulated by finite element models and simplified analytical formulations [23]. Satisfying agreement has been achieved in terms of global elastic properties (Young modulus and Poisson ratio) for almost all the studied topologies, including those exhibiting auxeticity along certain tested directions. The research findings have left some open issues regarding the tetrachiral material, exhibiting the highest discrepancy between numerical results and experimental evidences. In the same research field, successful comparisons between experiments and simulations on medium size samples have been also obtained for (i) rapidly prototyped chiral honeycombs in terms of transverse shear stiffness [24], (ii) laser-crafted re-entrant anti-trichiral honeycombs in terms of Young moduli and Poisson ratios [25], (iii) selective laser sintered tetrachiral and hexachiral honeycombs in terms of flatwise buckling loads [26]. The promising outcomes of experimental investigations on smart tetrachiral and hexachiral honeycomb have demonstrated the viability of advanced applications in the structural health monitoring and other sensing purposes [27].

More recently, honeycomb sheets with chiral and anti-chiral cellular topologies have been employed to develop three-dimensional curved structures. The auxeticity has been numerically simulated in microstructured cylindrical stents and experimentally verified in planar steel samples fabricated through the waterjet cutting technology [28]. Finite element results and experimental measures have been also compared to assess the coupled extensional and torsional deformations of microstructured cylindrical shells fabricated using selective laser sintering and selective laser melting methods with nylon and aluminium alloy [29]. Within this comparative study, the effects of a reinforcing rotation disk introduced at the cylinder mid-height have been also considered [30]. High-pressure abrasive waterjet technologies have been applied to realize functionally graded tetrachiral structures. The influence of the geometric gradient factor on the structural elastoplastic response has been numerically predicted and experimentally analysed during compressive load tests [31].

Negative Poisson ratios have been experimentally observed in hybrid chiral bi-dimensional materials fabricated via multi-material 3D printing. The auxetic behaviour, combined with sequential cell-opening mechanisms, has been purposely designed to develop innovative multi-functional composites, characterized by smart sensitivity to environmental conditions and targeted to technological applications in drug delivery and colour changing for camouflage [32]. New topologies thought for 3D auxetic material have been based on multi-layered tetrachiral schemes, with inter-layer clockwise and anticlockwise ligament connections. By tuning the direction of the interlayer connections, materials with two positive and one negative Poisson ratios have been first analytically predicted and successively confirmed through numerical simulations and experimental test on medium-size samples printed with the stereolithography technology [33]. Other different topologies of 3D chiral materials have been proposed and their deformation mechanisms have been experimentally studied through tensile and compression tests on selected laser sintered samples [34].

The most notable trend emerging from the review of the most recent state-of-the-art is the increasing and pervasive employment of additive manufacturing solutions for the engineering-oriented application of the large amount of theoretical knowledge about auxetic materials based on tailorable microstructures [35],[36],[37],[38]. Indeed, additive manufacturing is rapidly evolving as one of the most promising manufacturing technologies for designing, optimizing, rapid prototyping and large scale producing three-dimensional architected cellular materials with high-fidelity realization of complex microstructural topologies [39], [40], [41], [42], [43], [44]. Interesting advanced applications for additive manufactured architected materials range across many modern fields in frontier engineering, from micro-electro-mechanical systems to lightweight components for automotive or aerospace industry, from patient-specific medical implants to smart structural elements in parametric engineering and architecture.

The additive manufacturing process always starts from a virtual 3D model that has to be converted into a 3D printing-suitable format (the most common is the Standard Triangulation Language). Then, a so-called slicing procedure is performed and, for each slice, specific machine instructions are defined, which govern the 3D printer during the layer by layer production process. Among the others, one of the most widespread, versatile, and economic 3D printing processes is the Fused Deposition Modelling (FDM). This technology employs a thermoplastic material that is first heated to a semi-molten state and then extruded through a robotically-controlled nozzle in a temperature-controlled environment to construct layer by layer the desired part. Currently, a large variety of thermoplastic materials can be extruded, ranging from acrylonitrile butadiene styrene (ABS) and polylactic acid (PLA) to techno-polymers, like polyetherether ketone (Peek) or polytherimide (Ultem). Thanks to their superior mechanical properties, these thermoplastics can be used to produce also structurally functional components that, by virtue of the relatively low-cost of the technology, have been used for a wide spectrum of innovative technological applications ranging from acoustics and mechanics [45],[46],[47] to biomedicine and pharmaceutics [48],[49], from electronics [50],[52] to social applications [53],[54].

This stimulating and challenging scenario motivates the leading idea to conjugate the most recent progresses in additive manufacturing with the pressing demand to establish a robust experimental background supporting the most advanced theoretical and applied researches in the exotic elasticity and smart engineering functionality of existing and new architectured materials. According to these basic motivations, the paper leverages the actual technological possibility in realizing high-fidelity complex topologies by 3D printing thermoplastic materials in order to bridge the scientific gap between analytical or numerical predictions and experimental evidences in the field of chiral microstructured materials. Focus is laid on experimentally validating some theoretical results about the directional auxeticity of the tetrachiral material [25],[55],[56]. In this respect, planar polymeric samples have been 3D printed by employing the FMD technology (Section 1), in order to realize a tetrachiral cellular geometry with the highest possible precision (Section 1.1). Following a multidisciplinary approach for the data acquisition and processing, the samples have been tensile tested and the quasi-static response has been measured by means of non-contact technologies based on digital image acquisition (Section 1.2). Therefore, the measures have been analysed by virtue of numerical data post-processing to solve the inverse problem concerned with the input-output identification of the global elastic properties (Section 1.3). In parallel, different mechanical models of the tetrachiral samples have been developed (Section 2), in the framework of solid mechanics (Section 2.1) and structural mechanics (Section 2.2). Analytical and numerical solutions have been determined to simulate the experimental tests and compare the respective findings in terms of global Young modulus and Poisson ratio (Section 2.3). As valuable point of novelty, the research outcomes have led to the theoretical conceptualization, mechanical modelization and analytical/numerical simulation of an original by-layered topology, based on tetrachiral layers. The new topology, which differs from other layered auxetic materials based on radially foldable microstructure [57] and does not require a different bi-layered multi-material 3D printing process, is kinematically based on the independent and opposite-sign rolling up mechanisms of the component layers, reciprocally constrained at the boundaries. The theoretical predictions and the experimental behaviour have been compared in terms of global rigidity and auxeticity (Section 3). Finally, concluding remarks have been pointed out and future developments have been outlined.

1. TETRACHIRAL SAMPLE

The class of chiral and antichiral cellular materials is characterized by a periodic tessellation of the bidimensional plane. The elementary cell is strongly characterized by a microstructure composed by stiff circular rings connected by flexible straight ligaments, arranged according to different planar geometries including the trichiral, hexachiral, tetrachiral, anti-trichiral, anti-tetrachiral topologies. Among the others, the tetrachiral material is featured by a monoatomic centrosymmetric cell in which the central stiff and massive ring (or disk) is connected to four tangent flexible and light ligaments.

1.1 Sample preparation

A polymeric sample of the tetrachiral geometry has been realized with Fused Deposition Modelling (FDM) technology by the Group of Computational Mechanics and Advanced Materials of the University of Pavia. For the layer-by-layer FDM preparation of the samples a thermoplastic filament made of the polymer Acrylonitrile Butadiene Styrene (ABS) has been used. This material is known to have a Young modulus ranging in the large interval 1100-2900 *MPa* [58]. In the initial pre-print conditions, the nominal value given by the filament manufacturer is close to 2000 *MPa*. The reduced value for the post-print conditions is a matter of experimental identification and has been tentatively fixed at 1300 *MPa* as initial realistic value. The Poisson ratio has been realistically fixed at the nominal value of 0.35.

The printing head movements and all the printing parameters are automatically controlled by an electronic board relying on a set of instructions (i.e., the G-Code). The G-Code is produced by a dedicated software, commonly called *slicer* or *slicing software* that takes into account the virtual geometry, the properties of the printing material, and the specific features of the 3D-printer. The 3D printer used to produce the tetrachiral sample was a 3NTR A4v3 (see Figure 1a). The machine has been equipped with three extruders that can be heated up to 410 °C; a nozzle of 0.4 *mm* of diameter has been used. The build-tray temperature has been set to 110 °C, while the heated chamber temperature to 70 °C in order to avoid distortions of the printed sample induced by the high thermal gradients occurring during the manufacturing process. A filament cross-section with thickness of 0.2 *mm* and width of 0.4 *mm*, and a fiber-to-fiber overlap of 0.04 *mm* has been assumed. In Figure 1b the printed sample is shown. The geometric dimensions of the printed sample are reported in Figure 1c and detailed in Table 2.

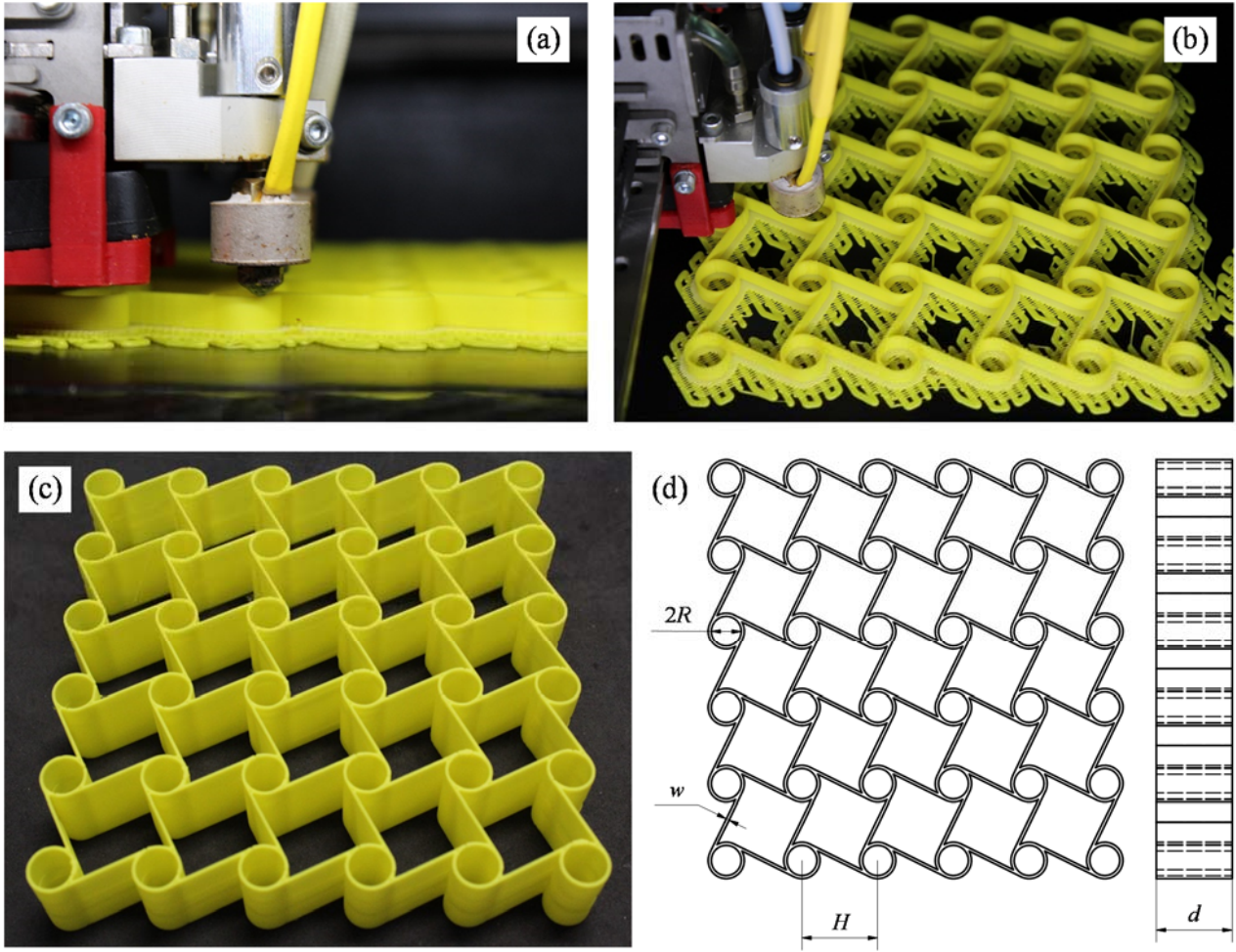

**Figure 1**. Tetrachiral sample printed with the FDM technology: (a), (b) 3NTR 3D printer preparing the tetrachiral sample; (c) 3D printed sample; (d) geometry of the printed sample.

Table 1. Geometric properties of the tetrachiral and bi-tetrachiral samples

|  | Dimensions (in *mm*) | Notes |
|---|---|---|
| Inter-ring distance $H$ | 20 | - |
| Ring mean radius $R$ | 4 | - |
| Ring width $w$ | 0.8 | - |
| Ligament width $w$ | 0.8 | - |
| Sample depth $d$ | 20 | Tetrachiral sample |
| Sample depth $d$ | 6 | Layer of the bi-tetrachiral sample |

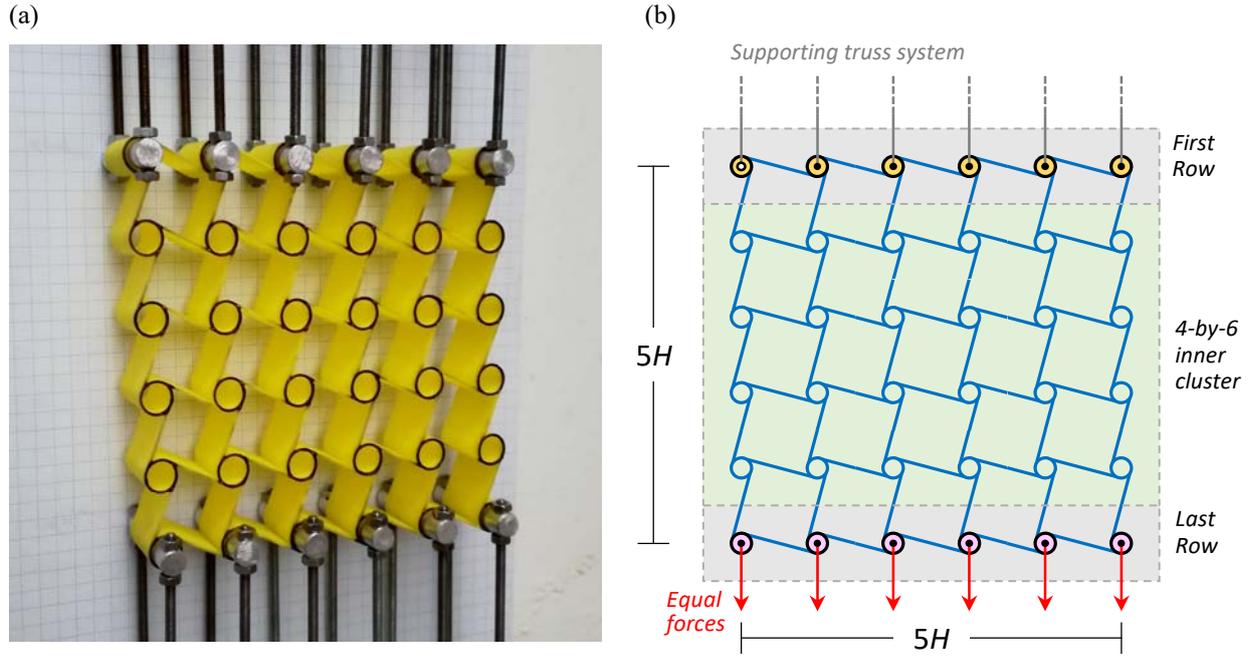

**Figure 2.** Experimental set up in the initial reference configuration: (a) picture, (b) sketch of the supported tetrachiral sample under the action of the external forces.

## 1.2 Experimental tests

The experimental activities have been carried out at the Laboratory of Structural Engineering of the DICCA – University of Genova. The tetrachiral samples made of an array of 6-by-6 cells have been tested under uni-axial tension according to a force-control scheme, within stable environmental conditions (Figure 2a). The test set up has been designed to apply known increments of force to the sample, starting from an initial reference configuration.

In the reference configuration, the sample is hanging vertically under self-weight load and supported by a constraining system applied at the top side (Figure 2b). The constraining system has been purposely designed to ideally preserve the alignment of all the rings located in the first (top) row of the sample (yellow-filled black rings in Figure 2b). The transversal displacements and rotations of all the rings remain ideally free, allowing the unrestricted development of both lateral expansion/contraction and the rolling-up mechanism. The controlled force acts at the bottom side of the sample by means of a steel truss system conveniently designed to split the total vertical action into equal forces (red vectors in Figure 2b) applied at all the rings located in the last (bottom) row of the sample (pink-filled black rings in Figure 2b). In order to assess the global elastic properties of the tetrachiral sample, the 4-by-6 inner cluster of internal unconstrained and unloaded cells (green region in Figure 2b) been considered in the following.

The uni-axial tension test has been run by applying five increments of the quasi-static force $F_2$, corresponding to equivalent global stress $\Sigma_{22} = F_2/A_2$, where $A_2 = 6Hd$ is the cross-section area of the ideal solid (rectangular parallelepiped) with dimensions $6H$ (width), $6H$ (height) and $d$ (depth). The force increments (steps 1-5) are reported in Table 2, starting from the initial loading conditions (step 0), under the self-weight of the sample and the truss system. Particular attention has been payed to some operational issues, like preserving the vertical planarity of the deformed configurations and minimizing the parasitic effects of friction in the constraints. Finally, the entire loading process has been designed not to overcome the limit of linear reversible deformation in the material of the sample. This design requirement has been checked a posteriori by verifying that the initial undeformed configuration is recovered at the end of the unloading process, here not reported for the sake of synthesis.

Table 2. Loading steps for the uni-axial tension test of the tetrachiral sample

|  | Force $F_2$ [N] | Global stress $\Sigma_{22}$ [N/m$^2$] |
|---|---|---|
| Step 0 | 8.15 | 4075 |
| Step 1 | 18.15 | 9075 |
| Step 2 | 28.15 | 14075 |
| Step 3 | 38.15 | 19075 |
| Step 4 | 48.15 | 24075 |
| Step 5 | 58.15 | 29075 |

## 1.3 Data processing and identification

### 1.3.1 *Image acquisition and post-processing*

Among the different available possibilities for data recording in quasi-static experimental tests, the non-contact data acquisition through a digital camera has been selected as a convenient compromise between measure reliability and operational feasibility [59]. Specifically, one or more two-dimensional images (with dimensions 5184-by-3456 pixels) of the tetrachiral specimen have been acquired for each load step with a fixed camera (14 bit Canon EOS 600D, with image processor DIGIC 4). In order to minimize distortion, the focal plane of the digital camera has been initially calibrated to be parallel to the specimen plane, with the focal axis crossing the geometric center of the specimen in the undeformed configuration.

All the digital photographs have been uploaded and converted in two-dimensional arrays of pixel coordinates within the Matlab environment [60], in order to be post-processed by using the Image Processing Toolbox. First, the background grid of known dimensions (approximately co-planar with the specimen) has been employed to convert the pixel coordinates of the digital photograph into a real coordinate system. Second, the real positions **x** of the ring centers have been assessed by programming an automatic function for the recognition of assigned shapes, based on the Hough transform. This mathematical transform is an efficient tool for geometric shape recognition, largely used in digital image processing and computer vision. The automatic recognition of an assigned shape is based on searching object imperfections within a certain class of geometric objects, in which the most suited candidates are recognized as local maxima in a finite parameter space (also known as accumulator space). Specifically, the Hough transform is classically concerned with the identification of lines in digital images, but it can be easily extended to identify positions of arbitrary curves. In particular, the recognition of circumferences is commonly used to identify the centers and radii of circles [61].

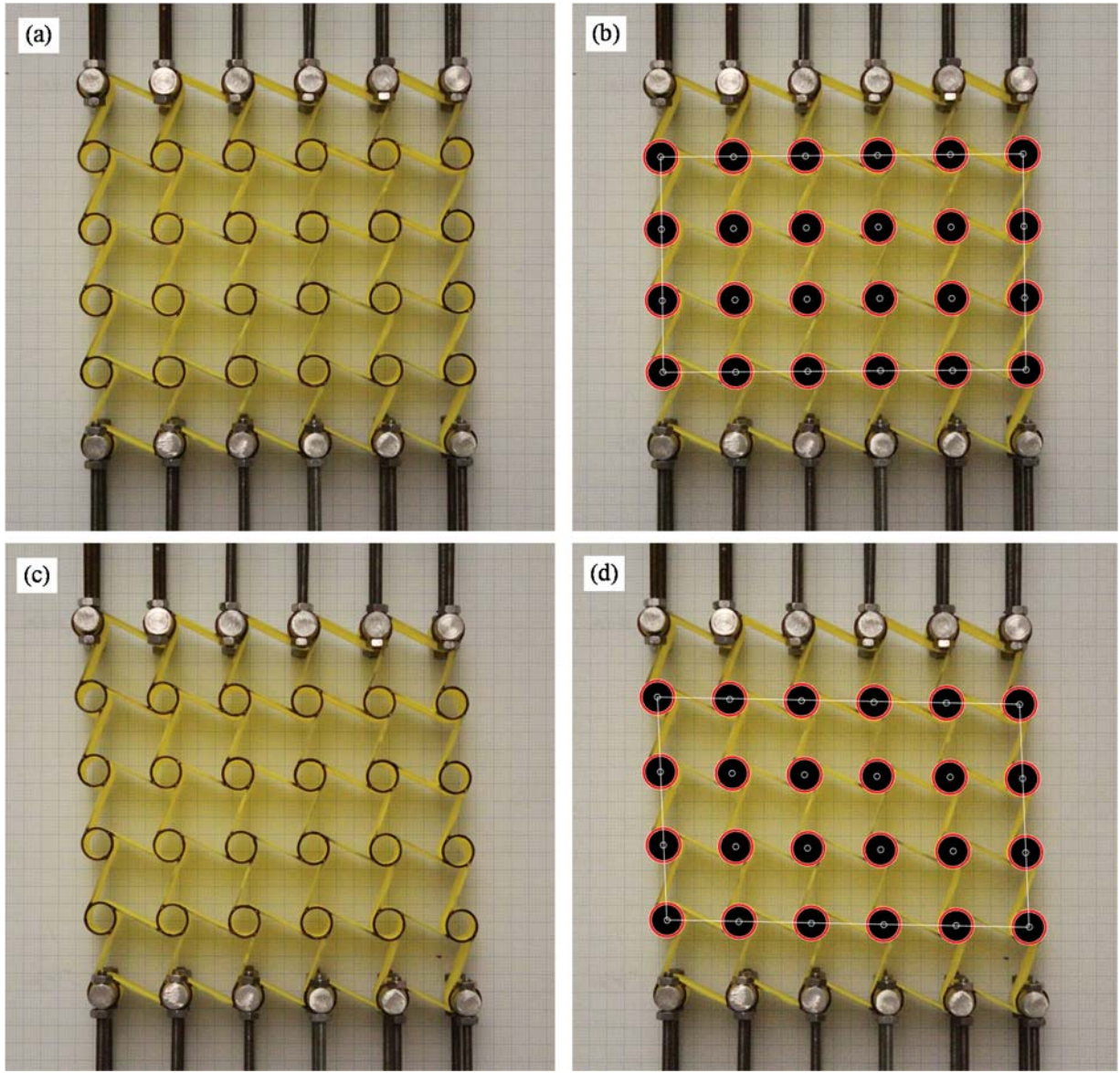

**Figure 3**. Non-contact image-based identification of the real coordinates for all the ring centers in the inner cluster of the tetrachiral sample: (a),(b) loading step 0; (c),(d) loading step 5.

By properly tuning the sensitivity parameters of the automatic function, the identification of the real coordinates of all the unconstrained ring centers in the tetrachiral specimen has been successfully run for each load step (Figure 3). Therefore, the total displacements of the ring

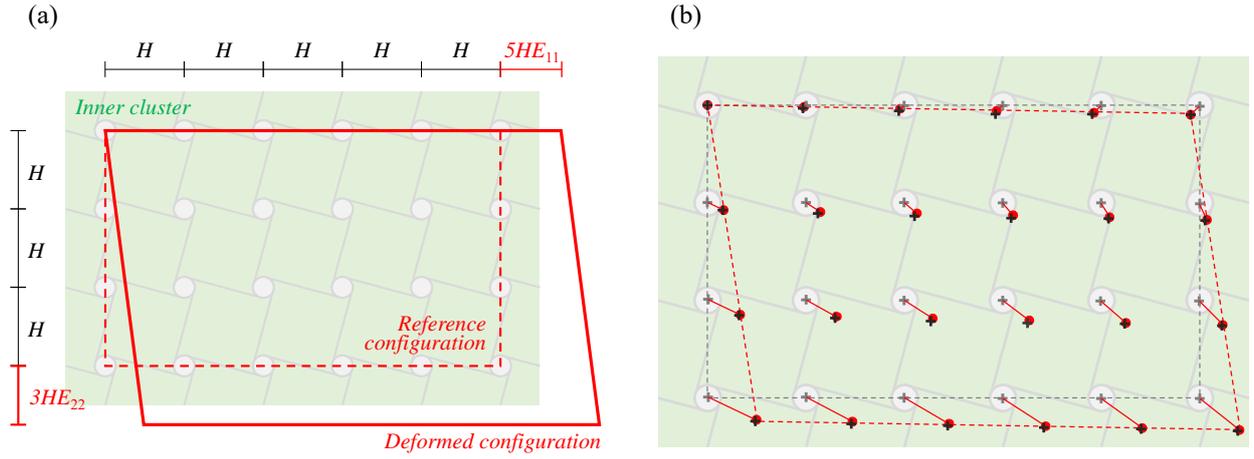

**Figure 4**. Identification of the global elastic properties: (a) affine displacement field and physical meaning of the normal strains $E_{11}$, $E_{22}$ related to the deformed configuration; (b) comparison between the experimental deformed configuration at the loading step 5 (black crosses) and the corresponding identified deformed configuration (red circles and red displacement vectors)

centers have been calculated as position differences with respect to the initial load step, and the results have been stored in the experimental set of punctual displacement vector $\mathbf{v}$.

*1.3.2 Input-output identification of the global elastic properties*

According to a mechanical model-based identification approach, the global elastic properties of the tetrachiral samples are identified by introducing the in-plane affine displacement field $\mathbf{u}$ of an equivalent homogeneous Cauchy continuum in the framework of a linear kinematics (Figure 4a). Accordingly, the continuum displacement field $\mathbf{u}(\mathbf{x})$ of the material point at the generic position $\mathbf{x}$ is expressed through the classic relation

$$\mathbf{u}(\mathbf{x}) = \mathbf{u}_0 + \mathbf{H}\mathbf{x} \qquad (1)$$

where $\mathbf{u}_0$ is the displacement of an arbitrary origin point. The two-by-two $\mathbf{H}$-matrix is the displacement gradient, composed by a skew-symmetric part $\mathbf{W}$ and a symmetric part $\mathbf{E}$ that can be recognized as infinitesimal rotation matrix and infinitesimal strain matrix, respectively.

The experimental set of punctual displacement vector $\mathbf{v}$ measured at known positions $\mathbf{x}$ can conveniently be supposed – as first approximation – to obey the linear vector law

$$\mathbf{v}(\mathbf{x}) = \mathbf{v}_0 + \mathbf{G}\mathbf{x} \qquad (2)$$

where $\mathbf{v}_0$ is the known displacement of a measurement point taken as origin and the two-by-two coefficient matrix $\mathbf{G}$ of linear proportionality is unknown a priori. Since the displacement variable $\mathbf{v}$ is experimentally known only in a finite number $m$ of measurement points (Figure 4b), the identification issue consists in imposing $m$ relations

$$\mathbf{v}_h = \mathbf{v}_0 + \mathbf{G}\mathbf{x}_h, \qquad h = 1,...,m \qquad (3)$$

and searching for the four independent $\mathbf{G}$-components. Once the $\mathbf{G}$-components have been determined, the symmetric and skew-symmetric parts of the $\mathbf{G}$-matrix can be extracted.

Finally, by recognizing the formal analogy between the equations (1) and (2), the two infinitesimal rotation and infinitesimal strain matrices can be univocally identified

$$\begin{aligned}\mathbf{E} &= \tfrac{1}{2}\left(\mathbf{H} + \mathbf{H}^T\right) = \tfrac{1}{2}\left(\mathbf{G} + \mathbf{G}^T\right) \\ \mathbf{W} &= \tfrac{1}{2}\left(\mathbf{H} - \mathbf{H}^T\right) = \tfrac{1}{2}\left(\mathbf{G} - \mathbf{G}^T\right)\end{aligned} \qquad (4)$$

Therefore, once the matrices $\mathbf{E}$ and $\mathbf{W}$ are identified, the normal strains $E_{11}$, $E_{22}$, the angular strain $E_{12}$ and the infinitesimal rotation $W_{21} = -W_{12}$ read

$$E_{11} = G_{11}, \quad E_{22} = G_{22}, \quad E_{12} = \tfrac{1}{2}(G_{12} + G_{21}), \quad W_{12} = \tfrac{1}{2}(G_{12} - G_{21}) \qquad (5)$$

Furthermore, the global elastic properties can be assessed by the well-known relations

$$E = \frac{\Sigma_{22}}{H_{22}}, \qquad \nu = -\frac{H_{11}}{H_{22}} \qquad (6)$$

where $E$ and $\nu$ are the global Young modulus and global Poisson ratio, respectively.

From the mathematical viewpoint, the identification problem can be properly formulated by stating the algebraic linear equation

$$\mathbf{A}\mathbf{y} = \mathbf{b} \qquad (7)$$

where $\mathbf{y} = (G_{11}, G_{22}, G_{12}, G_{21})$ is the four-by-one vector of unknowns, $\mathbf{A}$ is a known $2m$-by-four matrix depending only on the positions of the measurement points and $\mathbf{b}$ is a known $2m$-by-one depending only on the measured displacements.

In the common operational case corresponding to redundancy of measures ($m > 3$), the algebraic problem turns out to be overdetermined. Therefore, it is necessary to solve equation (7) according to the least square approximation of the solution, which reads

$$\mathbf{y} = \mathbf{A}^+\mathbf{b} \qquad (8)$$

where $\mathbf{A}^+$ stands for the pseudoinverse of the rectangular matrix $\mathbf{A}$.

## 2. NUMERICAL SIMULATIONS AND ANALYTICAL PREDICTIONS

### 2.1 Solid finite element model

The tetrachiral samples can be modeled according to a high-dimensional formulation in the framework of linear solid mechanics. The software Autodesk Inventor [62] has been used as CAD modeling tool to describe with high-fidelity the complex three-dimensional geometry of the generic tetrachiral samples composed by a bidimensional array of $N \times M$ cells, with $N$ rows and $M$ columns (Figure 5a illustrates the six-by-six cell sample). Therefore, the software

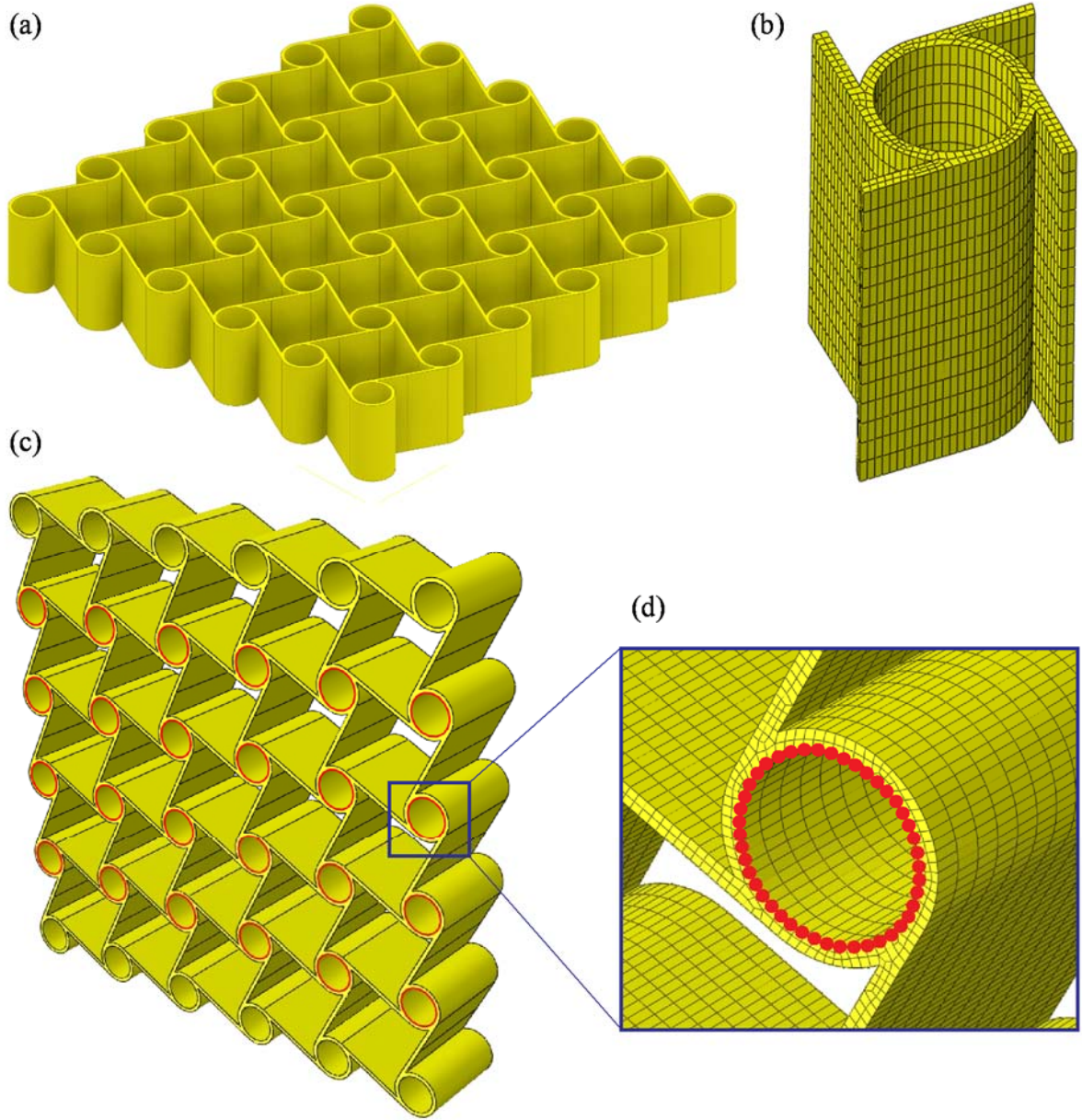

**Figure 5**. Three-dimensional solid model of the tetrachiral sample: (a) perspective geometric view of the six-by-six cell sample; (b) detail of the finite element mesh for the periodic cell, (c),(d) selection of finite element nodes for the displacement reconstruction of the ring centers in the inner cluster.

Abaqus Standard [63] has been used as finite element mesh generator and solver. The entire sample domain has been discretized with 8-node linear brick, reduced integration, hourglass control (C3D8R) elements. The six-by-six cell sample has been meshed in 146220 brick elements and the mesh detail of the periodic cell are illustrated in Figure 5b.

The solid model has been specified by assuming isotropic homogeneous material with elastic properties described by the Young modulus $E$ and the Poisson ratio $\nu$. To reproduce the operational conditions of the experimental tests, rigid body definitions have been applied to the nodes belonging to the internal surface of all the rings located in the first (top) row and the last (bottom) row of the sample. The boundary conditions at the top side of the tetrachiral sample have been imposed by applying external constraints at the rigid bodies of the first (top) cell row. Specifically, all the vertical displacements (plus one horizontal displacement to avoid rigid motion solutions) are constrained. Equal vertical forces have been applied to each centroid of all the rigid bodies of the last (bottom) cell row.

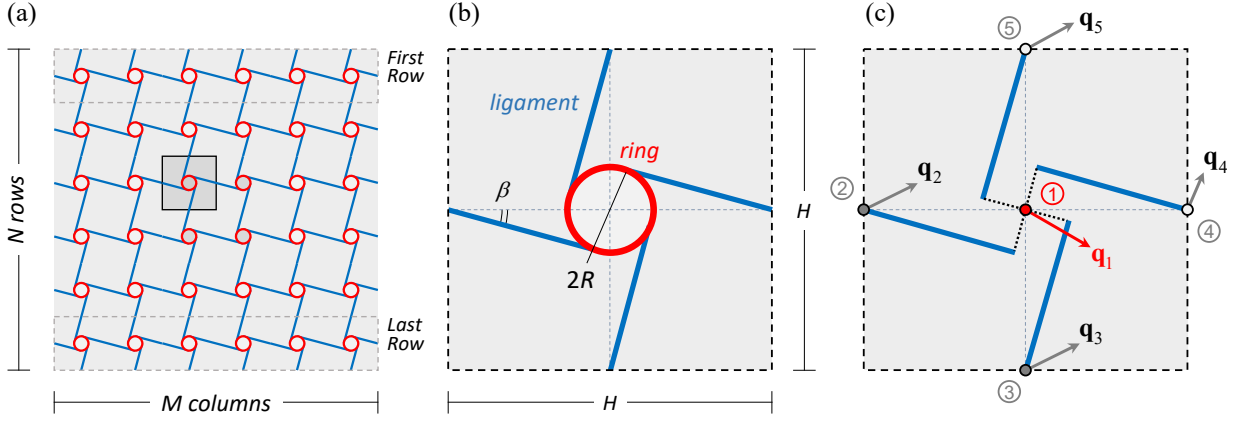

**Figure 6**. Tetrachiral sample: (a) repetitive planar pattern, (b) periodic cell, (c) beam lattice model.

From the finite element solution, the 48-by-one displacement subvector **v** that collects the horizontal and vertical displacement of the ring centers in the inner cluster of 4-by-6 cells has been reconstructed and subsequently employed to identify the global elastic properties. Consistently with the experimental measurements, the displacement reconstruction for all the ring centers in the inner cluster has been achieved by properly averaging the displacements of all the finite element nodes at the internal ring circumference (see red nodes in Figure 5c,d).

## 2.2 Beam lattice model

The periodic cell of the tetrachiral samples can be modelled according to a low-dimensional lagrangian formulation in the framework of linear mechanics (Figure 6a). The tetrachiral planar geometry is characterized by the side length $H$ of the periodic square cell and the mean radius $R$ of the rings (Figure 6b). According to simple trigonometric considerations, the chirality angle of the tangent ligaments is $\beta = \arctan(2R/H)$ and the ligament length is $L = H \cos \beta$.

The lagrangian model is synthesized on the base of a few mechanical assumptions. The stiff rings are conveniently described as annular rigid bodies, while the flexible ligaments are described as unshearable beams. A linear elastic material with Young modulus $E$ is assumed for all beams. Therefore, each beam is characterized by axial rigidity $EA$ and in-plane flexural rigidity $EJ$. Moreover, the beam-annulus connections are assumed perfectly rigid joints.

The deformed configuration of the periodic cell is fully described by three planar degrees-of-freedom for each configurational node. An internal configurational node is located at the annulus centroid, while four external configurational nodes are located at the beam midspan, where the cell boundary crosses the ligaments (Figure 6c). The degrees-of-freedom of the internal node and $i$-th external node (with $i=1,...,4$) are collected column-wise in the generalized displacement vector $\mathbf{q}_1 = (u_1, v_1, \vartheta_1)$ and $\mathbf{q}_i = (u_i, v_i, \vartheta_i)$, respectively, where $u$ stands for the rightward horizontal displacement, $v$ for the upward vertical displacement, $\vartheta$ for the counter-clockwise in-plane rotation. All the nodal degrees-of-freedom are collected in the 15-by-1 cellular displacement vector $\mathbf{q} = (\mathbf{q}_1, \mathbf{q}_2, \mathbf{q}_3, \mathbf{q}_4, \mathbf{q}_5)$. Following the direct stiffness method, the 15-by-15 stiffness matrix $\mathbf{K}$ of the generic cell can be determined. The non-zero submatrices of the stiffness matrix $\mathbf{K}$ are reported in Appendix A.

Considering the tetrachiral sample composed by a bidimensional array of $N \times M$ cells (with $N$ rows and $M$ columns), the global $3P$-by-one displacement vector $\mathbf{q}_g$ can be defined, where the total number of configurational nodes in the unconstrained lagrangian model is $P = 3NM + N - M$. Introducing the cellular-to-global change-of-coordinates $\mathbf{q}_j = \mathbf{\Lambda}_j \mathbf{q}_g$ for the $j$-th cell (with $j = 1,...,NM$), the $3P$-by-$3P$ global stiffness matrix reads

$$\mathbf{K}_g = \sum_{j=1}^{NM} \mathbf{\Lambda}_j^{\mathrm{T}} \mathbf{K} \mathbf{\Lambda}_j \qquad (9)$$

where $\mathbf{\Lambda}_j$ is the 15-by-$3P$ boolean allocation matrix. Consequently, the algebraic equations governing the static equilibrium of the unconstrained lagrangian model are

$$\mathbf{K}_g \mathbf{q}_g = \mathbf{f}_g \qquad (10)$$

where $\mathbf{f}_g$ is the $3P$-by-one vector of the forces acting in the components of the vector $\mathbf{q}_g$.

In order to describe the boundary conditions at the top side of the tetrachiral sample, opportune external constraints of the central nodes of the first (top) cell row must be imposed. Defining the $(M+1)$-by-one displacement vector $\mathbf{q}_c$ that collects all the vertical displacements (plus one horizontal displacement to avoid rigid motion solutions) of the central nodes in the $M$ cells of the first (top) row, the following partition can be introduced

$$\begin{bmatrix} \mathbf{K}_{ff} & \mathbf{K}_{fc} \\ \mathbf{K}_{cf} & \mathbf{K}_{cc} \end{bmatrix} \begin{pmatrix} \mathbf{q}_f \\ \mathbf{q}_c \end{pmatrix} = \begin{pmatrix} \mathbf{f}_f \\ \mathbf{f}_c \end{pmatrix} \qquad (11)$$

where the $(3P - M - 1)$-by-one vector $\mathbf{q}_f$ collects all the free degrees-of-freedom, playing the role of lagrangian coordinates. Indeed, imposing the constraint condition $\mathbf{q}_c = \mathbf{0}$, the equation (11) can be inverted to obtain the static equilibrium solution

$$\begin{aligned} \mathbf{q}_f &= \mathbf{K}_{ff}^{-1} \mathbf{f}_f \\ \mathbf{f}_c &= \mathbf{K}_{cf} \mathbf{K}_{ff}^{-1} \mathbf{f}_f \end{aligned} \qquad (12)$$

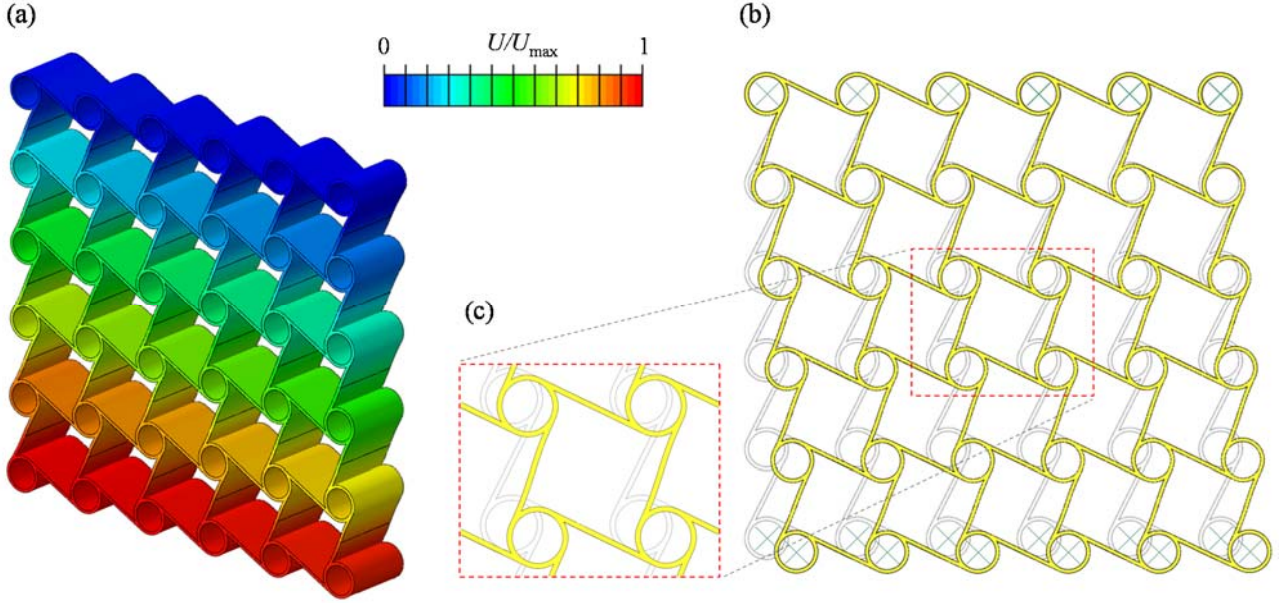

**Figure 7**. Finite element simulation of the tetrachiral sample: (a) color map of the horizontal displacement component; (b),(c) comparison between the undeformed (gray lines) and deformed configuration (yellow solid).

where $\mathbf{f}_f$ has non-null equal components corresponding to the only vertical forces acting in the central nodes of the last (bottom) cell row. In order to extract from the displacement solution $\mathbf{q}_f$ the pseudo-experimental results necessary for the identification of the global elastic properties, the 48-by-one displacement subvector $\mathbf{v}$ that collects the horizontal and vertical displacement of the central nodes in the inner cluster of 4-by-6 cells is considered.

## 2.3 Comparison of experiments and simulations

The experimental results are compared qualitatively and quantitatively with the simulations obtained with the lagrangian model and with the finite element solid model of the tetrachiral sample. First, from a qualitative point of view, all the experimental deformed configurations (see for instance Figure 3c) and the corresponding static simulations agree in exhibiting a non-symmetric behaviour in response to the application of the symmetric scheme (with respect to the vertical in-plane axis crossing the sample barycenter) of external forces. This behaviour can be essentially attributed to the chirality of the cellular topology, which determines the development of local deformation mechanisms, activated by the rotation of the rings around their centers (*rolling up*). Globally, the vertical stretching of the sample – collinear to the external force direction – is accompanied with an evident and significant angular strain. This characteristic behaviour is observable in the deformed configurations obtained from the experimental test (Figure 3d), as well as in the numerical simulations obtained from the solid model (Figure 7). It is worth noting that the sign of the angular strain is univocally related to the sign of the chirality angle in the cellular topology (angle $\beta$ in Figure 6b). For the numerical

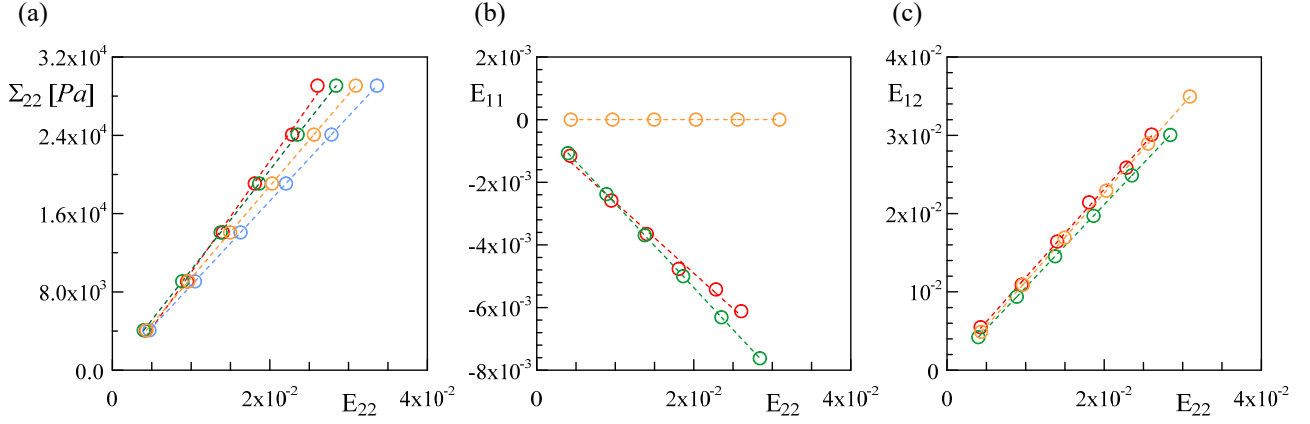

**Figure 8**. Comparison between experimental results (red circles) and numerical results of the solid model (blue and green circles) and beam lattice model (yellow circles): (a) equivalent global stress $\Sigma_{22}$ versus normal strain $E_{22}$, (b) normal strain $E_{11}$ versus normal strain $E_{22}$, (c) angular strain $E_{12}$ versus normal strain $E_{22}$

simulations, the horizontal component of the normalized displacement field $U/U_{max}$ is shown in Figure 7a, while the deformed and undeformed configurations are compared in Figure 7b,c.

Second, from a quantitative point of view, the normal and angular strains identified starting from the experimental data are compared with the corresponding strains identified starting from the numerical results obtained with the Lagrangian and the finite element solid models, respectively (Figure 8). As a major remark, it can be noted that the experimental response (red circles) shows – with a good approximation – a linear behaviour under increasing values of the external forces. With focus on the solid model, it can be observed that the ratio $\Sigma_{22}/E_{22}$ obtained from the numerical results (blue circles in Figure 8a) systematically underestimates the corresponding experimental ratio (red circles in Figure 8a). This occurrence has demonstrated the need to update the nominal value initially assumed for the Young modulus of the ABS material (1300 *MPa*), which has been recognized as the most uncertain mechanical parameter according to the initial information available. Therefore, this parameter has been properly updated in order to zeroing the difference between the global Young modulus $E$ identified from the numerical results (load-independent value marked by green circles in Figure 9a) and the average of the global Young modules ($E=1.02\,MPa$), identified from the experimental results at different loading steps (red circles in Figure 9a). The updating procedure has required an increment in the Young modulus of the ABS material of about 18% (up to 1540 *MPa*). The ratio $\Sigma_{22}/E_{22}$ obtained from the numerical results of the updated solid model (green circles in Figure 8a) shows a better agreement with the corresponding experimental ratio (red circles in Figure 8a).

Adopting the updated solid model, a very satisfactory agreement is achieved between the ratio $E_{11}/E_{22}$ obtained from the numerical results (green circles or, rigorously, the slope of the dashed line connecting the green circles in Figure 8b) and the corresponding experimental ratio (red circles or, rigorously, slope of the dashed line linearly regressing the red circles in Figure 8b). The negative value systematically attained by this ratio for increasing external forces demonstrates a non-auxetic behaviour of the tetrachiral sample in the direction orthogonal to the external forces. This key result is effectively synthesized by the positive values systematically identified for the global Poisson ratio $\nu$ (Figure 9b). In particular, the load-independent value ($\nu = 0.27$) identified from the numerical results (green circles in Figure 9b) closely matches the mean value ($\nu = 0.26$) identified from the experimental results (red circles in Figure 9b). It is worth noting that these two consistent values are also in good agreement with the analytical results obtained from a second gradient continuum model of the tetrachiral material formulated according to a proper homogenization technique [55]. Finally, the first qualitative remark concerned with the development of a non-negligible angular strain is quantitatively confirmed by the identification of the strain $E_{12}$ that assumes experimental and numerical values in mutual agreement and quantitatively comparable with the values of the normal strain $E_{22}$ (green and red circles in Figure 8c).

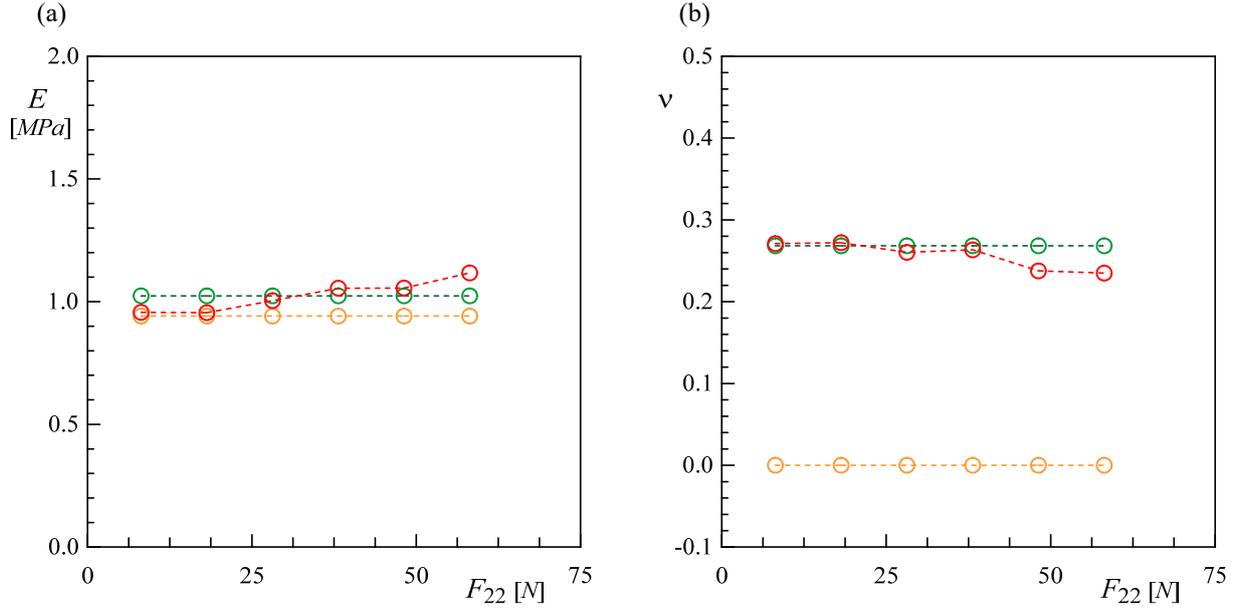

**Figure 9.** Comparison between experimental results (red circles) and numerical results of the solid model (green circles) and beam lattice model (yellow circles). Global elastic properties versus external load $F_{22}$: (a) Young modulus $E$, (b) Poisson ratio $\nu$.

With focus on the Lagrangian model, a good matching is found in the simulation (yellow circles in Figure 8a) of the experimental ratio $\Sigma_{22}/E_{22}$. Differently, the global Young modulus identified from the Lagrangian simulation (load-independent value marked by yellow circles in Figure 9a) returns a slight underestimation ($E = 0.94\,MPa$) of the mean experimental value (red circles in Figure 9a). This difference is a well-established finding that can be attributed to the rigid body assumption for the central ring and to the overestimation of the effective length in the flexible ligaments composing the cellular microstructure [55]. Furthermore, the simplifying assumptions of the Lagrangian formulation do not allow the accurate assessment of the experimental value for the ratio $E_{11}/E_{22}$ (compare red and yellow circles in Figure 8b). Indeed, it can be highlighted that the corresponding load-independent identification of the global Poisson ratio ($\nu \approx 0$ marked by yellow circles in Figure 9b) is perfectly consistent with the null value ($\nu = 0$) exactly obtainable by a micropolar continuum model of the tetrachiral material formulated according to a proper continualization technique [55]. In contrast, the Lagrangian model accurately captures the positive experimental ratio $E_{12}/E_{22}$ (yellow circles in Figure 8c), which measures the coupling between the normal strain $E_{22}$ and the angular strain $E_{12}$.

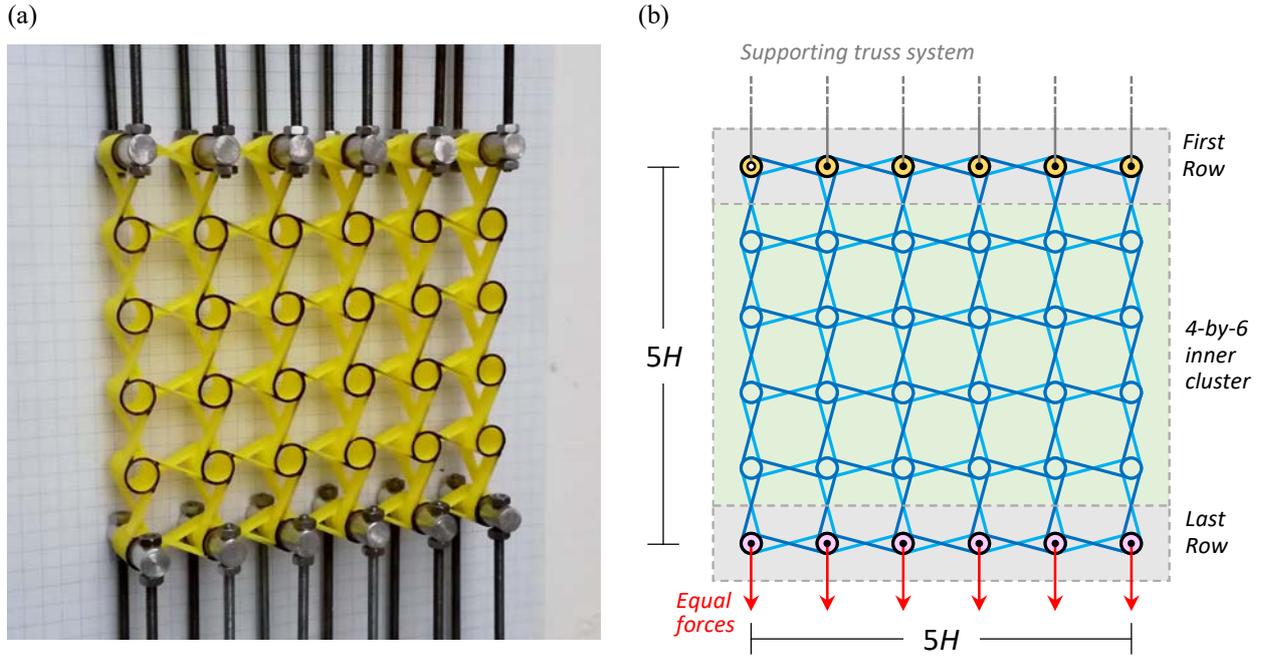

**Figure 10.** Experimental set up in the initial reference configuration: (a) picture, (b) sketch of the supported bi-tetrachiral sample under the action of the external forces.

## 3. BI-TETRACHIRAL SAMPLE

### 3.1 Sample preparation, Experimental set-up, Data acquisition and processing

The bi-layered polymeric sample of the bi-tetrachiral geometry has been obtained by simply superimposing two tetrachiral samples (front and back layers) with opposite chirality. The geometric dimensions of the printed sample are reported in Table 2.

As for the tetrachiral samples, the bi-tetrachiral samples made of an array of 6-by-6 cells have been experimentally tested under uni-axial tension according to the force-control scheme, within stable environmental conditions. Known increments of force have been applied to the sample, starting from the initial constrained configuration (Figure 10a), taken as reference. In the reference configuration, the sample lies in the vertical plane under self-weight load and is supported by the alignment-preserving constraining system. The supporting system is designed to constrain the relative displacements (but not the relative rotation) of all the 6 ring pairs (one ring in the front layer, the other in the back layer) at the top side. The steel truss system acting at the bottom side of the bi-tetrachiral sample applies the controlled force and simultaneously constrains the relative displacements (but not the relative rotation) of all the 6 ring pairs (one ring in the front layer, the other in the back layer) at the bottom side. The total vertical action is split into equal forces applied at all the ring pairs located in the last (bottom) row of the sample (Figure 10b). Apart from the relative constraints at the top and bottom sides, each layer is free to independently develop both lateral expansion/contraction and the rolling-up mechanism. In order to assess the global elastic properties of the bi-tetrachiral sample, the 4-by-6 inner cluster of internal unconstrained and unloaded cells has been considered in the following.

The uni-axial tension test has been run by applying five increments of the quasi-static force $F_2$, corresponding to equivalent global stress $\Sigma_{22} = F_2/A_2$, where $A_2 = 12Hd$ is the cross-section area of the ideal solid with dimensions $6H$ (width), $6H$ (height) and $2d$ (depth). The force increments (steps 1-5) are the same already reported in Table 2, starting from the initial configuration (step 0). Attention has been paid to preserve the vertical planarity of the deformed configurations, minimizing as much as possible the undesired effects of friction in the constraints and not overcoming the limit of linear reversible deformation in the material.

As for the tetrachiral samples, the non-contact data acquisition has been performed through a digital camera. Specifically, one or more two-dimensional images (with dimensions 5184-by-3456 pixels) of the bi-tetrachiral specimen have been acquired for each load step, taking care of minimizing distortion by properly calibrating the focal axis and focal plane of the camera according to the specimen plane and geometry. After uploading the digital photographs in the Matlab environment, the Image Processing Toolbox has been employed to convert the pixel coordinates of the digital photograph into a real coordinate system. The real positions **x** of the ring centers of the layers have been again assessed by recognizing the circular shapes by virtue of the Hough transform for each load step. Therefore, the total displacement vector **v** of the ring centers have been calculated as position differences with respect to the initial load step. Finally, the global elastic properties (Young modulus $E$ and Poisson ratio $\nu$) are assessed following the procedure described in Section 1.3.2.

## 3.2 Numerical simulations and analytical predictions

### 3.2.1 Solid finite element model

In analogy with the tetrachiral samples, the bi-tetrachiral samples can be modelled according to a high-dimensional formulation in the framework of linear solid mechanics. The two layers have been first reproduced with high geometrical fidelity in the framework of the CAD tool [62] (Figure 11a illustrates the six-by-six cell sample), and successively discretized and analyzed using Abaqus [63]. The same mesh size and the element type of the tetrachiral solid model have been adopted for the two layers of the bi-tetrachiral samples (for instance, the six-by-six cell sample has been meshed in 129834 C3D8R elements and the mesh detail of the periodic cell are illustrated in Figure 11b).

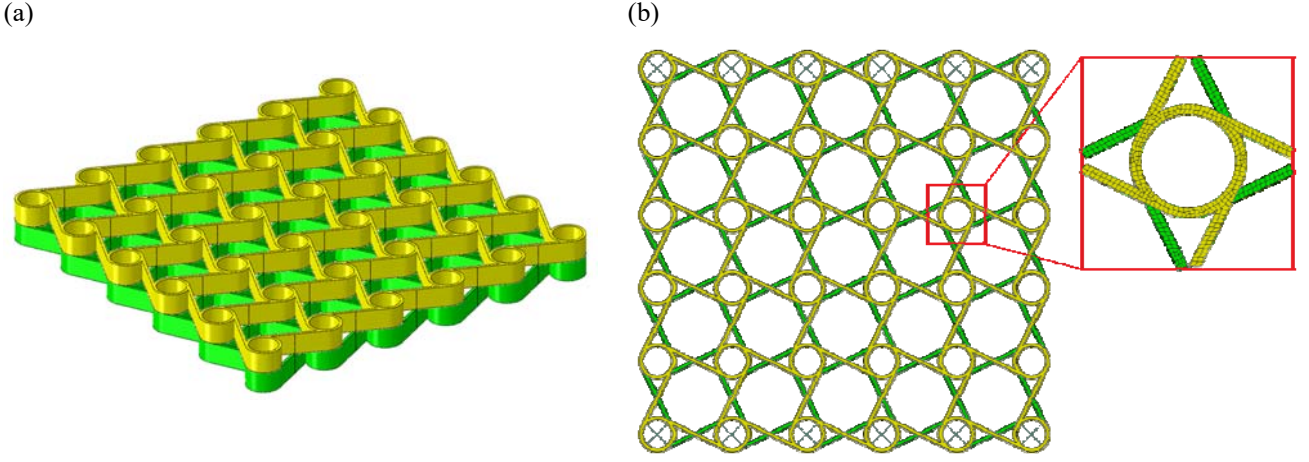

**Figure 11**. Three-dimensional solid model of the bi-tetrachiral sample: (a) perspective geometric view of the six-by-six cell sample; (b) finite element mesh and detail of the periodic cell.

The solid model has been specified by assuming the same isotropic homogeneous material, the same rigid body assumptions and the same external boundary conditions of the tetrachiral model. In addition, inter-layer constraints have been defined to impose the cell-to-cell identity between: (i) the pairs of horizontal displacements of the ring centers in the first (top) row of the two layers, (ii) the pairs of horizontal displacements and the pairs of vertical displacements of the ring centers in the last (bottom) row of the two layers. Equal vertical forces have been applied to each centroid of all the rigid bodies of the last (bottom) cell row in the front and back layer. From the finite element solution, the 48-by-one displacement subvector $\mathbf{v}$ that collects the horizontal and vertical displacement components of the ring centers in the inner cluster of 4-by-6 cells has been reconstructed and then employed to identify the equivalent elastic parameters.

### 3.2.2 Beam lattice model

In analogy with the tetrachiral samples, the bi-layered periodic cell of the bi-tetrachiral samples can again be modelled according to a low-dimensional lagrangian formulation (Figure 12a). The independent geometric properties of the periodic square cell are the side length $H$ and the ring radius $R$, while the chirality angles of the tangent ligaments are $\beta = \pm \arctan(2R/H)$, where the positive and the negative signs correspond to the front layer (identified by the superscript (1) in the following) and to the back one (superscript (2)), respectively (Figure 12b).

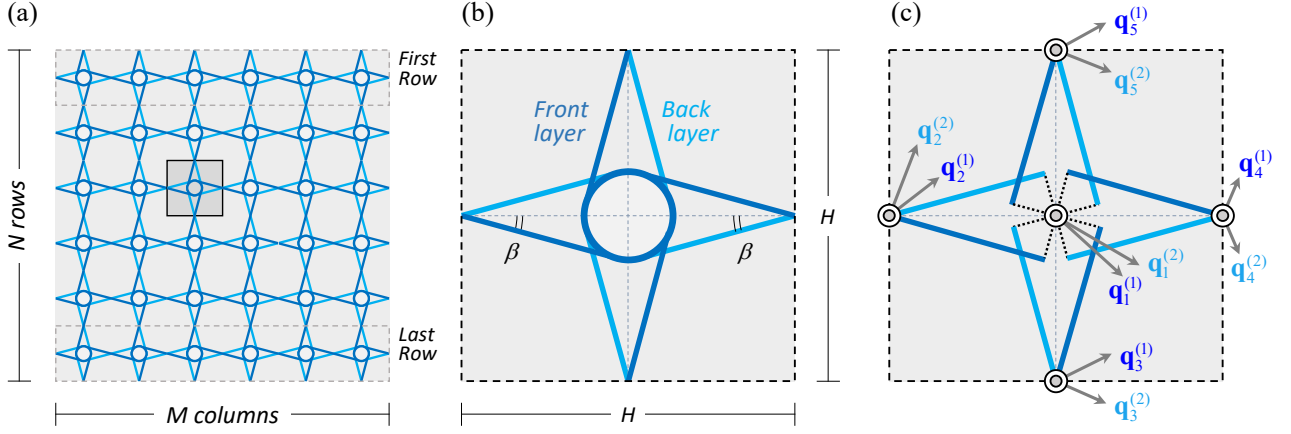

**Figure 12**. Bi-Tetrachiral sample: (a) repetitive planar pattern, (b) periodic cell, (c) beam lattice model.

Based on the simplifying mechanical assumptions already introduced for the lagrangian model of the tetrachiral samples, identical structural properties (stiff rings, flexible ligaments with Young modulus $E$, axial and flexural rigidities $EA$ and $EJ$) are considered for the two layers of the bi-tetrachiral samples. In the reference planar configuration, the two internal nodes (one for each layer) and the eight external nodes (four for each layer) share the same position.

The deformed configuration of the periodic cell is fully described by three planar degrees-of-freedom for each configurational node (Figure 12c). The degrees-of-freedom of the internal node and the $i$-th external node (with $i = 1,...,4$) of the $k$-th layer are collected column-wise in the two vectors $\mathbf{q}_1^{(k)} = \left(u_1^{(k)}, v_1^{(k)}, \vartheta_1^{(k)}\right)$ and eight vectors $\mathbf{q}_i^{(k)} = \left(u_i^{(k)}, v_i^{(k)}, \vartheta_i^{(k)}\right)$, respectively. All the nodal degrees-of-freedom are collected in the two 15-by-1 cellular displacement vectors $\mathbf{q}^{(1)} = \left(\mathbf{q}_1^{(1)}, \mathbf{q}_2^{(1)}, \mathbf{q}_3^{(1)}, \mathbf{q}_4^{(1)}, \mathbf{q}_5^{(1)}\right)$ and $\mathbf{q}^{(2)} = \left(\mathbf{q}_1^{(2)}, \mathbf{q}_2^{(2)}, \mathbf{q}_3^{(2)}, \mathbf{q}_4^{(2)}, \mathbf{q}_5^{(2)}\right)$. The 15-by-15 stiffness matrix $\mathbf{K}^{(1)}$ of the generic cell in the front layer is the same already presented for the tetrachiral periodic cell in Appendix A, while the 15-by-15 stiffness matrix $\mathbf{K}^{(2)}$ of the generic cell in the back layer is obtained straightforwardly by changing the $\beta$-sign.

Considering a bidimensional array of $N \times M$ cells (with $N$ rows and $M$ columns), the global $3P$-by-one displacement vectors $\mathbf{q}_g^{(1)}$ and $\mathbf{q}_g^{(2)}$ can be defined for two layers of the bi-tetrachiral samples. Introducing the cellular-to-global change-of-coordinates $\mathbf{q}_j^{(k)} = \mathbf{\Lambda}_j \mathbf{q}_g^{(k)}$ for the $j$-th cell (with $j = 1,...,NM$), the $3P$-by-$3P$ global stiffness matrices read

$$\mathbf{K}_g^{(1)} = \sum_{j=1}^{NM} \mathbf{\Lambda}_j^{\mathrm{T}} \mathbf{K}^{(1)} \mathbf{\Lambda}_j, \qquad \mathbf{K}_g^{(2)} = \sum_{j=1}^{NM} \mathbf{\Lambda}_j^{\mathrm{T}} \mathbf{K}^{(2)} \mathbf{\Lambda}_j \qquad (13)$$

where the 15-by-$3P$ boolean allocation matrices $\mathbf{\Lambda}_j$ hold for both the layers. Joining the layer displacement vectors into the $6P$-by-1 global displacement vector $\mathbf{q}_g = \left(\mathbf{q}_g^{(1)}, \mathbf{q}_g^{(2)}\right)$, the $6P$-by-$6P$ stiffness matrix of the bi-tetrachiral sample reads

$$\mathbf{K}_g = \begin{bmatrix} \mathbf{K}_g^{(1)} & \mathbf{O} \\ \mathbf{O} & \mathbf{K}_g^{(2)} \end{bmatrix} \qquad (14)$$

Consequently, apart the doubled dimension, the algebraic equations governing the static equilibrium of the unconstrained lagrangian model are formally identical to the equations of the tetrachiral samples

$$\mathbf{K}_g \mathbf{q}_g = \mathbf{f}_g \tag{15}$$

where $\mathbf{f}_g$ is the $6P$-by-one vector of the forces acting in the components of the vector $\mathbf{q}_g$.

In order to describe the boundary conditions at the top side of the bi-tetrachiral sample, proper external constraints of the central nodes of the first (top) cell row must be imposed. Defining the $2(M+1)$-by-one displacement vector $\mathbf{q}_c$ that collects all the vertical displacements (plus a pair of horizontal displacements, one for each layer, to avoid rigid motion solutions) of the central nodes in the $2M$ cells of the first (top) row, the following partition can be introduced

$$\begin{bmatrix} \mathbf{K}_{uu} & \mathbf{K}_{uc} \\ \mathbf{K}_{cu} & \mathbf{K}_{cc} \end{bmatrix} \begin{pmatrix} \mathbf{q}_u \\ \mathbf{q}_c \end{pmatrix} = \begin{pmatrix} \mathbf{f}_u \\ \mathbf{f}_c \end{pmatrix} \tag{16}$$

where the $2(3P-M-1)$-by-one vector $\mathbf{q}_u$ collects all the externally unconstrained degrees-of-freedom. Internal constrains are also necessary to describe the inter-layer coupling at the top and bottom boundaries. To this purpose, the unconstrained displacement vector can be decomposed as $\mathbf{q}_u = (\mathbf{q}_m, \mathbf{q}_s)$, where the $6(P-M)$-by-one master displacement vector $\mathbf{q}_m$ and the $2(2M-1)$-by-one slave displacement vector $\mathbf{q}_s$ are distinguished. Therefore, the internal constraints can be imposed on the slave vector through the relation $\mathbf{q}_s = \mathbf{V}\mathbf{q}_m$, where the $2(2M-1)$-by-$6(P-M)$ boolean constraint matrix $\mathbf{V}$ imposes the cell-to-cell identity between: (i) the pairs of horizontal displacements of the central nodes in the first (top) row of the two layers, (ii) the pairs of horizontal displacements and the pairs of vertical displacements of the central nodes in the last (bottom) row of the two layers. Accordingly, the force vector can be decomposed $\mathbf{f}_u = (\mathbf{f}_m, \mathbf{f}_s)$. Consequently, the equation (16) can be partitioned in the convenient form

$$\begin{bmatrix} \mathbf{K}_{mm} & \mathbf{K}_{ms} & \mathbf{K}_{mc} \\ \mathbf{K}_{sm} & \mathbf{K}_{ss} & \mathbf{K}_{sc} \\ \mathbf{K}_{cm} & \mathbf{K}_{cs} & \mathbf{K}_{cc} \end{bmatrix} \begin{pmatrix} \mathbf{q}_m \\ \mathbf{q}_s \\ \mathbf{q}_c \end{pmatrix} = \begin{pmatrix} \mathbf{f}_m \\ \mathbf{f}_s \\ \mathbf{f}_c \end{pmatrix} \tag{17}$$

and, imposing the constraint condition $\mathbf{q}_c = \mathbf{0}$, the equation (17) can be inverted to obtain the static equilibrium solution

$$\begin{aligned} \mathbf{q}_m &= (\mathbf{K}_{mm} + \mathbf{K}_{ms}\mathbf{V})^{-1} \mathbf{f}_m \\ \mathbf{f}_s &= (\mathbf{K}_{sm} + \mathbf{K}_{ss}\mathbf{V})(\mathbf{K}_{mm} + \mathbf{K}_{ms}\mathbf{V})^{-1} \mathbf{f}_m \\ \mathbf{f}_c &= (\mathbf{K}_{cm} + \mathbf{K}_{cs}\mathbf{V})(\mathbf{K}_{mm} + \mathbf{K}_{ms}\mathbf{V})^{-1} \mathbf{f}_m \end{aligned} \tag{18}$$

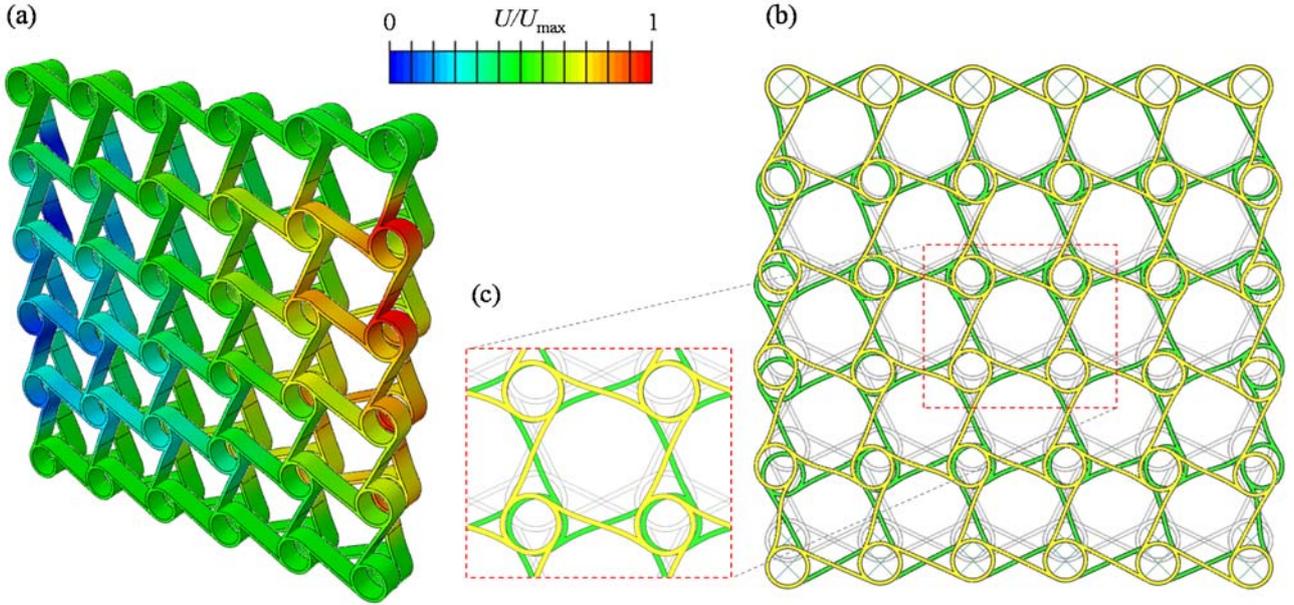

**Figure 13**. Finite element simulation of the bi-tetrachiral sample: (a) color map of the horizontal displacement component; (b),(c) comparison between the undeformed (gray lines) and deformed configuration (yellow and green solids).

where $\mathbf{f}_m$ has non-null equal components corresponding to the only vertical forces of the central nodes in the last (bottom) row of the two layers. Again, in order to extract from the displacement solution $\mathbf{q}_m$ the pseudo-experimental results necessary for the identification of the equivalent elastic parameters, the 48-by-one displacement subvector $\mathbf{v}$ that collects the horizontal and vertical displacement of the central nodes in the inner cluster of 4-by-6 cells is considered.

### 3.3 Comparison of experiments and simulations

The experimental results are compared qualitatively and quantitatively with the simulations obtained with the lagrangian and the finite element solid model of the bi-tetrachiral sample. First, from a qualitative point of view, the two layers exhibit a different but mutually and doubly symmetric behaviour in all the experimental deformed configurations under the symmetric scheme of external forces. Specifically, the response of the two-layer sample is symmetric with respect to the vertical and horizontal in-plane axes crossing the sample barycenter. The double symmetry of the experimental response is confirmed by the static simulations. This peculiar behaviour can be essentially attributed to the opposite chiralities of the two layers composing the cellular topology. The opposite chirality angles determine the concurrent development of two independent mechanisms of local deformation, one for each layer, activated by the opposite-sign rotations of the rings around their centers. Globally, the vertical stretching of the sample – collinear to the external force direction – is not accompanied with any appreciable angular strain. This characteristic behaviour is observable in the experimental deformed configurations, as well as in the numerical simulations obtained from the solid model (Figure 13).

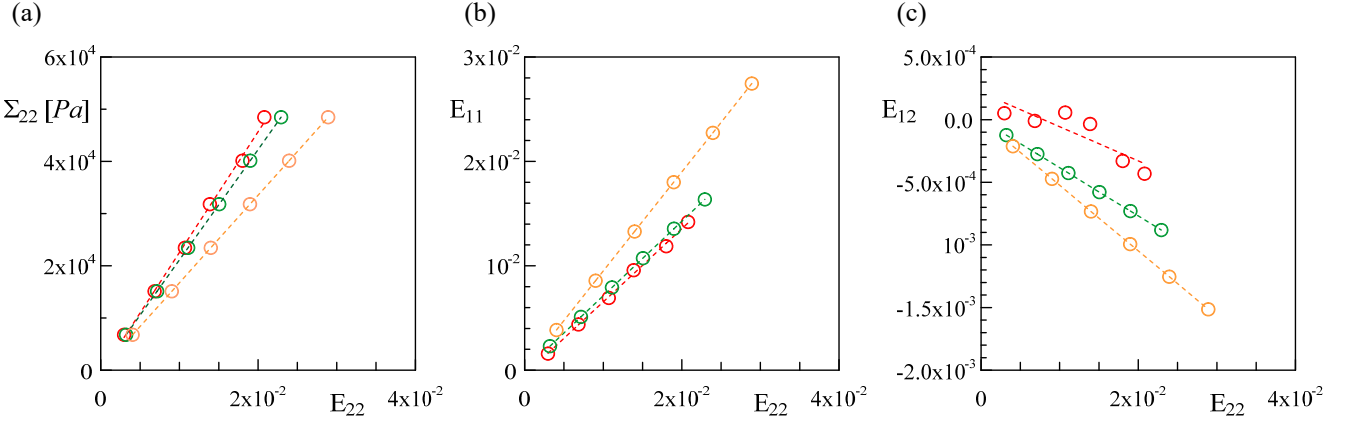

**Figure 14**. Comparison between experimental results (red circles) and numerical results of the solid model (green circles) and beam lattice model (yellow circles): (a) equivalent global stress $\Sigma_{22}$ versus normal strain $E_{22}$, (b) normal strain $E_{11}$ versus normal strain $E_{22}$, (c) angular strain $E_{12}$ versus normal strain $E_{22}$

The horizontal component of the normalized displacement field $U/U_{max}$ is shown in Figure 13a, while the deformed configurations of the front layer (yellow) and back layer (green) are portrayed in Figure 13b,c. Second, from a quantitative point of view, the normal and angular strains identified from the experimental data are compared with the corresponding strains identified from the numerical results obtained with the lagrangian and the finite element solid models (Figure 14), respectively. As for the tetrachiral sample, it can be noted that the experimental response (red circles) shows a quasi-perfectly linear behaviour under increasing values of the external forces.

With focus on the solid model, which employs the updated Young modulus of the ABS material (1540 $MPa$), the ratio $\Sigma_{22}/E_{22}$ obtainable from the numerical results of the updated solid model (green circles in Figure 14a) shows a fine agreement – apart from a minor underestimation – with the corresponding experimental ratio (red circles in Figure 14a). Consistently with this remark, the global Young modulus ($E = 2.11 MPa$) identified from the numerical results (green circles in Figure 15a) is slightly lower than the minimum and the mean value ($E = 2.26 MPa$) of the global Young modules identified from the experimental results (circles reds in Figure 15a). In synthesis, it can be highlighted how the global Young modulus of the bi-tetrachiral sample – in virtue of the two layer collaboration – is about twice that of the tetrachiral sample made of the same material.

Adopting the updated solid model, a very satisfactory agreement is again achieved between the ratio $E_{11}/E_{22}$ obtained from the numerical results (green circles in Figure 14b) and the corresponding experimental ratio (red circles in Figure 14b). However, the positive values systematically attained by this ratio for increasing external forces demonstrate a remarkably auxetic behaviour of the bi-tetrachiral sample in the direction orthogonal to the external forces. This key result fundamentally differentiates the global behaviour of the bi-tetrachiral material from that of the tetrachiral material individually characterizing the two component layers. From a mechanical perspective, this drastic change in the static response can be fully attributed to the inter-layer rigid constraint at the boundaries (top and bottom cell rows). The auxetic behaviour is effectively synthesized by the negative values systematically identified for the global Poisson ratio $\nu$ (Figure 15b). In particular, the load-independent value ($\nu = -0.71$) identified from the numerical results (green circles in Figure 15b) effectively matches – apart from a small underestimation – the mean value ($\nu = -0.64$) identified from the experimental results (red circles in Figure 15b). Finally, the first qualitative remark concerned with the absence of an appreciable angular strain is quantitatively confirmed by the identification of the angular strain $E_{12}$, which assumes experimental and numerical values in mutual agreement but significantly lower (by two orders of magnitude) than those of the normal strain $E_{22}$ (green and red circles in Figure 14c).

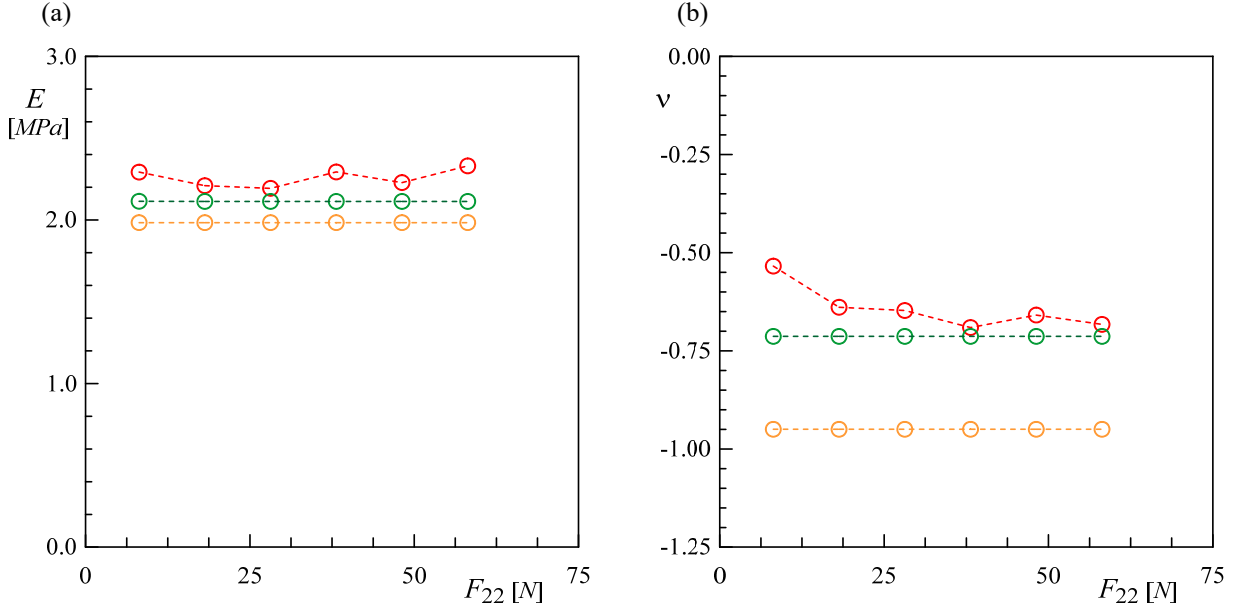

**Figure 15**. Comparison between experimental results (red circles) and numerical results of the solid model (green circles) and beam lattice model (yellow circles). Global elastic properties versus external load $F_{22}$: (a) Young modulus $E$, (b) Poisson ratio $\nu$.

With focus on the lagrangian model, a satisfying agreement is found in the simulation (yellow circles in Figure 14a) of the experimental ratio $\Sigma_{22}/E_{22}$. The identification of the global Young modulus from the lagrangian simulations (load-independent values marked by yellow circles in Figure 15a) returns a slight underestimation ($E = 1.98\,MPa$) of the corresponding values identified from the experiments (red circles) and the finite element solid model (green circles). Similarly to the tetrachiral sample, this difference can be attributed to the simplifying assumptions of the lagrangian formulation [55]. Despite the simplifying assumptions, the lagrangian formulation allows the identification of positive values for the ratio $E_{11}/E_{22}$, consistently with the experimental results. Nonetheless, it can be noted that the corresponding load-independent identification of the global Poisson ratio ($\nu = -0.95$ marked by yellow circles in Figure 15b) systematically overestimates the actual auxeticity of the bi-tetrachiral sample. Finally, the lagrangian model confirms a certain accuracy in fitting the experimental ratio $E_{12}/E_{22}$ (yellow circles in Figure 14c).

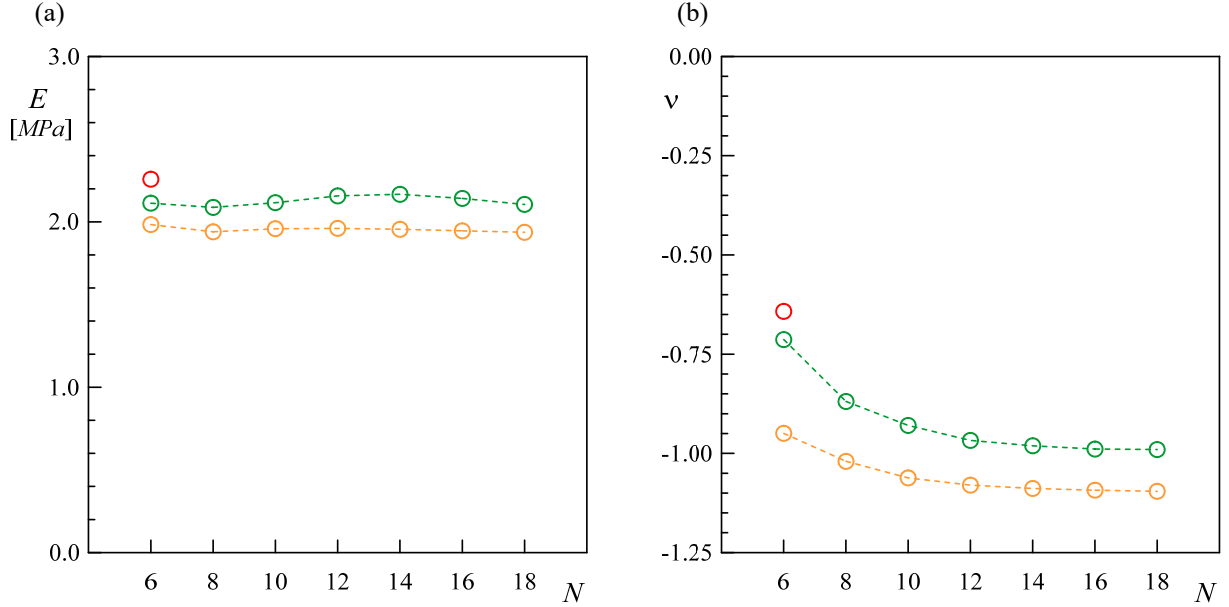

**Figure 16**. Comparison between experimental results (red circle) and numerical results of the solid model (green circles) and beam lattice model (yellow circles). Global elastic properties versus cell number of the square sample (with $N=M$): (a) Young modulus $E$, (b) Poisson ratio $\nu$.

Once the actual reliability of the mechanical formulations in simulating the static response of the bi-tetrachiral material has been experimentally verified, the solid and the lagrangian models have been used to simulate the behaviour of square samples characterized by increasing size (up to $N=M=18$). Given the key role played by the inter-layer boundary constraints on the auxetic response of the bi-tetrachiral material, the purpose of these simulations is targeted to verifying the influence of the boundary-to-boundary distance on the global elastic properties. Regardless of the sample size, the identification of the global Young modulus and global Poisson ratio has been based on the displacement subvector $\mathbf{v}$ related to the inner cluster of 4-by-6 cells, for the sake of consistency. The values of the global Young modulus identified from the solid model (green circles) and from the lagrangian model (yellow circles) exhibit limited variability for increasing sizes of the samples (Figure 16a). Specifically, $E$-values ranging from 2.08 *MPa* to 2.14 *MPa* have been obtained for the solid model (respectively for $N=8$ and $N=14$), while $E$-values ranging from 1.94 *MPa* and 1.98 *MPa* have been obtained for the lagrangian model (respectively for $N=8$ and $N=6$). Differently, the values of the global Poisson ratio identified from the solid model (green circles) and from the lagrangian model (yellow circles) exhibit a monotonically convergent behaviour for increasing sizes of the samples. For the largest size ($N=18$) the global Poisson ratio attains the minimum values $\nu=-0.99$ for the solid model and $\nu=-1.10$ for the lagrangian model (Figure 16b). This persistent result tends to confirm that the auxeticity is a property of the bi-tetrachiral material that actually leverages the inter-layer boundary constraints, but does not vanish for large boundary-to-boundary distances.

**CONCLUSIONS**

The global elastic properties of architectured honeycomb materials characterized by a tetrachiral cellular microstructure, made of circular stiff rings tangentially connected by flexible ligaments, have been investigated. The theoretical predictions based on low-dimension beam lattice models and finite element solid formulations have been compared with the experimental data obtained from quasi-static laboratory tests performed on different planar samples realized with the 3D printing technology. As major experimental evidence, the tetrachiral samples have shown a remarkable coupling between the normal and angular strains under increasing tension loads, due to the local ring rotations activated by the cellular chirality (*rolling up* mechanism). For the particular test direction, the global Young modulus and the global (positive) Poisson ratio of the tetrachiral material samples, experimentally identified by solving an overdetermined inverse problem, are accurately predicted by the numerical results of the solid computational model. On the contrary, some simplifying mechanical assumptions tend to limit the actual descriptive possibilities of the beam lattice model, particularly if the ring deformability is comparable with the ligament flexibility.

Inspired by the encouraging findings concerning the tetrachiral material, an original bi-layered topology of microstructured tetrachiral materials (*bi-tetrachiral materials*) has been theoretically conceived and mechanically modelled. The bi-tetrachiral topology exploits the virtuous mutual collaboration of two tetrachiral layers with opposite chirality in order to prevent the development of angular strains under tensile loads. As major technological advantage, the new material topology does not require a different bi-layered multi-material 3D printing process. The angular strain elimination is kinematically based on the independent and opposite-sign rolling up mechanisms of the two component layers, reciprocally constrained at the sample boundaries. Specifically, the boundary constraints are properly designed to impose identical layer-to-layer displacements to the rings, which remain free to rotate independently.

As main macroscopic consequence of its microstructural layout, the static response of the bi-tetrachiral material – composed of collaborating non-auxetic layers – exhibits a remarkably strong auxeticity that can be quantified by negative global Poisson ratio close to -0.7 for the tested samples. Moreover, the bi-tetrachiral material is found to outperform the tetrachiral material also in terms of global Young modulus, which turns out to be nearly doubled under the same testing conditions. These experimental results have been closely confirmed by the qualitative and quantitative simulations obtainable with beam lattice models and finite element solid formulations. In this case, the simplifying mechanical assumptions affecting the beam lattice model determine only a minor underestimation of the global Young modulus and the global Poisson ratio. Finally, parametric analyses have been carried out to evaluate numerically the effects of the bi-tetrachiral sample size on the global elastic properties. The result trend has demonstrated that a minimum Poisson ratio close to -1 can actually be reached by increasing the distance between the sample boundaries providing the interlayer constraint. This systematic results tend to confirm that the strong auxetic behaviour is substantially independent of the sample size, that is, represents a characteristic property of the bi-tetrachiral material.

**ACKNOWLEDGEMENTS**

The authors ML, FV and LG acknowledge financial support of the (MURST) Italian Department for University and Scientific and Technological Research in the framework of the research MIUR Prin15 project 2015LYYXA8 entitled *Multi-scale mechanical models for the design and optimization of micro-structured smart materials and metamaterials*, coordinated by prof. A. Corigliano. The author AB acknowledges financial support by National Group of Mathematical Physics (GNFM-INdAM). The authors FA and SM acknowledge the partial support by the strategic project of the University of Pavia entitled *Virtual Modeling and Additive Manufacturing for Advanced Materials* (3D@UniPV). Finally, all the authors acknowledge Dr. Gianluca Aliamo for technical support in printing all the samples, and Dr. Giuseppe Riotto for the assistance in designing and performing the experimental tests.

**COMPLIANCE WITH ETHICAL STANDARDS**

The authors declare that they have no conflict of interest

**REFERENCES**

[1] Fleck, N.A., Deshpande, V.S., Ashby, M.F. (2010). Micro-architectured materials: past, present and future. Proceedings of the Royal Society A: Mathematical, Physical and Engineering Sciences, 466(2121), 2495-2516.

[2] Schaedler, T.A., Carter, W.B. (2016). Architected cellular materials. Annual Review of Materials Research, 46, 187-210.

[3] Meza, L.R., Zelhofer, A.J., Clarke, N., Mateos, A.J., Kochmann, D.M., Greer, J.R. (2015). Resilient 3D hierarchical architected metamaterials. Proceedings of the National Academy of Sciences, 112(37), 11502-11507.

[4] Chen, Y.J., Scarpa, F., Liu, Y.J., Leng, J.S. (2013). Elasticity of anti-tetrachiral anisotropic lattices. International Journal of Solids and Structures, 50(6), 996-1004.

[5] Saxena, K.K., Das, R., Calius, E.P. (2016). Three decades of auxetics research − materials with negative Poisson's ratio: a review. Advanced Engineering Materials, 18(11), 1847-1870.

[6] Ha, C.S., Plesha, M.E., Lakes, R.S. (2016). Chiral three-dimensional isotropic lattices with negative Poisson's ratio. Physica status solidi (b), 253(7), 1243-1251.

[7] Lakes, R.S. (2017). Negative-Poisson's-Ratio Materials: Auxetic Solids. Annual Review of Materials Research, 47, 63-81.

[8] Ren, X., Das, R., Tran, P., Ngo, T.D., Xie, Y.M. (2018). Auxetic metamaterials and structures: A review. Smart materials and structures, 27(2), 023001.

[9] Duncan, O., Shepherd, T., Moroney, C., Foster, L., Venkatraman, P., Winwood, K., Allen T., Alderson, A. (2018). Review of auxetic materials for sports applications: Expanding options in comfort and protection. Applied Sciences, 8(6), 941.

[10] Spadoni, A., Ruzzene, M., Gonella, S., Scarpa, F. (2009). Phononic properties of hexagonal chiral lattices. Wave motion, 46(7), 435-450.

[11] Tee, K.F., Spadoni, A., Scarpa, F., Ruzzene, M. (2010). Wave propagation in auxetic tetrachiral honeycombs. Journal of Vibration and Acoustics, 132(3), 031007.

[12] Lepidi, M., Bacigalupo, A. (2018). Parametric design of the band structure for lattice materials. Meccanica, 53(3), 613-628.

[13] Vadalà, F., Bacigalupo, A., Lepidi, M., Gambarotta, L. (2018). Bloch wave filtering in tetrachiral materials via mechanical tuning. Composite Structures, 201, 340-351.

[14] Bacigalupo, A., Lepidi, M., Gnecco, G., Vadalà, F., Gambarotta, L. (2019). Optimal design of the band structure for beam lattice metamaterials. Frontiers in Materials, 6, 2.

[15] Bettini, P., Airoldi, A., Sala, G., Di Landro, L., Ruzzene, M., Spadoni, A. (2010). Composite chiral structures for morphing airfoils: Numerical analyses and development of a manufacturing process. Composites Part B: Engineering, 41(2), 133-147.

[16] Liu, X.N., Hu, G.K., Sun, C.T., Huang, G.L. (2011). Wave propagation characterization and design of two-dimensional elastic chiral metacomposite. Journal of Sound and Vibration, 330(11), 2536-2553.

[17] Ranjbar, M., Boldrin, L., Scarpa, F., Neild, S., Patsias, S. (2016). Vibroacoustic optimization of anti-tetrachiral and auxetic hexagonal sandwich panels with gradient geometry. Smart Materials and Structures, 25(5), 054012.

[18] Bacigalupo, A., Gambarotta, L. (2017). Wave propagation in non-centrosymmetric beam-lattices with lumped masses: Discrete and micropolar modeling. International Journal of Solids and Structures, 118, 128-145.

[19] Tallarico, D., Movchan, N.V., Movchan, A.B., Colquitt, D.J. (2017). Tilted resonators in a triangular elastic lattice: chirality, Bloch waves and negative refraction. Journal of the Mechanics and Physics of Solids, 103, 236-256.

[20] Bacigalupo, A., Lepidi, M. (2018). Acoustic wave polarization and energy flow in periodic beam lattice materials. International Journal of Solids and Structures, 147, 183-203.

[21] Lakes, R. (1991). Deformation mechanisms in negative Poisson's ratio materials: structural aspects. Journal of materials science, 26(9), 2287-2292.

[22] Prall, D., Lakes, R.S. (1997). Properties of a chiral honeycomb with a Poisson's ratio of -1. International Journal of Mechanical Sciences, 39(3), 305-314.

[23] Alderson, A., Alderson, K.L., Attard, D., Evans, K.E., Gatt, R., Grima, J.N., Miller, W., Ravirala, N., Smith, C.W., Zied, K. (2010). Elastic constants of 3-, 4-and 6-connected chiral and anti-chiral honeycombs subject to uniaxial in-plane loading. Composites Science and Technology 70(7), 1042-1048.

[24] Lorato, A., Innocenti, P., Scarpa, F., Alderson, A., Alderson, K.L., Zied, K.M., Ravirala N., Miller W., Smith C.W., Evans, K.E. (2010). The transverse elastic properties of chiral honeycombs. Composites Science and Technology 70(7), 1057-1063.

[25] Alderson, A., Alderson, K.L., Chirima, G., Ravirala, N., Zied, K.M. (2010). The in-plane linear elastic constants and out-of-plane bending of 3-coordinated ligament and cylinder-ligament honeycombs. Composites Science and Technology 70(7), 1034-1041.

[26] Miller, W., Smith, C.W., Scarpa, F., Evans, K.E. (2010). Flatwise buckling optimization of hexachiral and tetrachiral honeycombs. Composites Science and Technology 70(7), 1049-1056.

[27] Abramovitch, H., Burgard, M., Edery-Azulay, L., Evans, K.E., Hoffmeister, M., Miller, W., Scarpa F., Smith C.W., Tee, K.F. (2010). Smart tetrachiral and hexachiral honeycomb: Sensing and impact detection. Composites Science and Technology 70(7), 1072-1079.

[28] Li, H., Ma, Y., Wen, W., Wu, W., Lei, H., Fang, D. (2017). In plane mechanical properties of tetrachiral and antitetrachiral hybrid metastructures. Journal of Applied Mechanics, 84(8), 081006.

[29] Ma, C., Lei, H., Hua, J., Bai, Y., Liang, J., Fang, D. (2018). Experimental and simulation investigation of the reversible bi-directional twisting response of tetra-chiral cylindrical shells. Composite Structures, 203, 142-152.

[30] Wu, W., Geng, L., Niu, Y., Qi, D., Cui, X., Fang, D. (2018). Compression twist deformation of novel tetrachiral architected cylindrical tube inspired by towel gourd tendrils. Extreme Mechanics Letters, 20, 104-111.

[31] Li, M., Lu, X., Zhu, X., Su, X., Wu, T. (2018). Research on in-plane quasi-static mechanical properties of gradient tetra-chiral hyper-structures. Advanced Engineering Materials, 1801038, doi: 10.1002/adem.201801038.

[32] Jiang, Y., Li, Y. (2018). Novel 3d-printed hybrid auxetic mechanical metamaterial with chirality-induced sequential cell opening mechanisms. Advanced Engineering Materials 20(2), 1700744.

[33] Fu, M., Liu, F., Hu, L. (2018). A novel category of 3D chiral material with negative Poisson's ratio. Composites Science and Technology, 160, 111-118.

[34] Wu, W., Qi, D., Liao, H., Qian, G., Geng, L., Niu, Y., Liang, J. (2018). Deformation mechanism of innovative 3D chiral metamaterials. Scientific reports, 8(1), 12575.

[35] Guo, N., Leu, M.C. (2013). Additive manufacturing: technology, applications and research needs. Frontiers of Mechanical Engineering, 8(3), 215-243.

[36] Gao, W., Zhang, Y., Ramanujan, D., Ramani, K., Chen, Y., Williams, C.B., Wang C.C.L., Shin Y.C., Zhang S., Zavattieri, P.D. (2015). The status, challenges, and future of additive manufacturing in engineering. Computer-Aided Design, 69, 65-89.

[37] Jiang, Y., Li, Y. (2017). 3D printed chiral cellular solids with amplified auxetic effects due to elevated internal rotation. Advanced Engineering Materials, 19(2), 1600609.

[38] Chen, D., Zheng, X. (2018). Multi-material additive manufacturing of metamaterials with giant, tailorable negative Poisson's ratios. Scientific reports, 8, 9139.

[39] Clausen, A., Wang, F., Jensen, J.S., Sigmund, O., Lewis, J.A. (2015). Topology optimized architectures with programmable Poisson's ratio over large deformations. Advanced Materials, 27(37), 5523-5527.

[40] Schaedler, T.A., Carter, W. B. (2016). Architected cellular materials. Annual Review of Materials Research, 46, 187-210.

[41] Muth, J.T., Dixon, P.G., Woish, L., Gibson, L.J., Lewis, J.A. (2017). Architected cellular ceramics with tailored stiffness via direct foam writing. Proceedings of the National Academy of Sciences, 114(8), 1832-1837.

[42] Dong, G., Tang, Y., Zhao, Y.F. (2017). A survey of modeling of lattice structures fabricated by additive manufacturing. Journal of Mechanical Design, 139(10), 100906.

[43] Maskery, I., Sturm, L., Aremu, A.O., Panesar, A., Williams, C.B., Tuck, C.J., Wildman R.D., Ashcroft I.A., Hague, R.J.M. (2018). Insights into the mechanical properties of several triply periodic minimal surface lattice structures made by polymer additive manufacturing. Polymer, 152, 62-71.

[44] Vyatskikh, A., Delalande, S., Kudo, A., Zhang, X., Portela, C.M., Greer, J.R. (2018). Additive manufacturing of 3D nano-architected metals. Nature communications, 9(1), 593.

[45] Vayre, B., Vignat, F., Villeneuve, F. (2012). Metallic additive manufacturing: state-of-the-art review and prospects. Mechanics & Industry, 13(2), 89-96.

[46] Sha, Y., Jiani, L., Haoyu, C., Ritchie, R.O., Jun, X. (2018). Design and strengthening mechanisms in hierarchical architected materials processed using additive manufacturing. International Journal of Mechanical Sciences, 149, 150-163.

[47] Melde, K., Mark, A.G., Qiu, T., Fischer, P. (2016). Holograms for acoustics. Nature, 537(7621), 518.

[48] Auricchio, F., Marconi, S. (2016). 3D printing: clinical applications in orthopaedics and traumatology. EFORT open reviews, 1(5), 121-127.

[49] Capel, A.J., Rimington, R.P., Lewis, M.P., Christie, S.D. (2018). 3D printing for chemical, pharmaceutical and biological applications. Nature Reviews Chemistry, 2, 422-436.

[50] Macdonald, E., Salas, R., Espalin, D., Perez, M., Aguilera, E., Muse, D., Wicker, R.B. (2014). 3D printing for the rapid prototyping of structural electronics. IEEE access, 2, 234-242.

[51] Massoni, E., Silvestri, L., Bozzi, M., Perregrini, L., Alaimo, G., Marconi, S., Auricchio, F. (2016). Characterization of 3D-printed dielectric substrates with different infill for microwave applications. In 2016 IEEE MTT-S Int. Microwave Workshop Series on Advanced Materials and Processes for RF and THz Applications (IMWS-AMP).

[52] Cui, H., Hensleigh, R., Yao, D., Maurya, D., Kumar, P., Kang, M.G., Priya S., Zheng, X.R. (2019). Three-dimensional printing of piezoelectric materials with designed anisotropy and directional response. Nature materials, 18(3), 234-241.

[53] Huang, S.H., Liu, P., Mokasdar, A., Hou, L. (2013). Additive manufacturing and its societal impact: a literature review. The International Journal of Advanced Manufacturing Technology, 67(5-8), 1191-1203.

[54] Auricchio, F., Greco, A., Alaimo, G., Giacometti, V., Marconi, S., and Mauri, V. (2017). 3D printing technology for buildings accessibility: the tactile map for MTE museum in Pavia, Journal of Civil Engineering and Architecture, 11, 736-747.

[55] Bacigalupo, A., Gambarotta, L. (2014). Homogenization of periodic hexa-and tetrachiral cellular solids. Composite Structures, 116, 461-476.

[56] Chen, Y., Liu, X.N., Hu, G.K., Sun, Q.P., Zheng, Q.S. (2014). Micropolar continuum modelling of bi-dimensional tetrachiral lattices. Proceedings of the Royal Society A: Mathematical, Physical and Engineering Sciences, 470(2165), 20130734.

[57] Cabras, L., Brun, M. (2014). Auxetic two-dimensional lattices with Poisson's ratio arbitrarily close to -1. Proceedings of the Royal Society of London Series A, 470, 20140538-20140538.

[58] Coble, S. (2003) Materials data book. Cambridge University Engineering Department, Cambridge, UK.

[59] Gil, S., Reisin, H.D., Rodríguez, E.E. (2006). Using a digital camera as a measuring device. American journal of physics, 74(9), 768-775.

[60] MATLAB Full Suite (2019), The Math Works Inc., Natick, USA.

[61] Yuen, H.K., Princen, J., Illingworth, J., Kittler, J. (1990). Comparative study of Hough transform methods for circle finding. Image and vision computing, 8(1), 71-77.

[62] Autodesk Inventor (2017), Autodesk, San Rafael, USA.

[63] Abaqus (2017), Abaqus Theory Manual, SIMULIA - Dassault Systèmes, Providence, USA.

**APPENDICE A**

**A.1   Structural matrices**

The 15-by-15 stiffness matrix $\mathbf{K}$ of the periodic cell defined in Section 2.2 for the tetrachiral sample can be expressed in the form

$$\mathbf{K} = \begin{bmatrix} \mathbf{K}_{11} & \mathbf{K}_{12} & \mathbf{K}_{13} & \mathbf{K}_{14} & \mathbf{K}_{15} \\ & \mathbf{K}_{22} & \mathbf{O} & \mathbf{O} & \mathbf{O} \\ \ldots & & \mathbf{K}_{33} & \mathbf{O} & \mathbf{O} \\ & & & \mathbf{K}_{44} & \mathbf{O} \\ \text{Sym} & & \ldots & & \mathbf{K}_{55} \end{bmatrix} \quad (19)$$

where the three-by-three non-null submatrices are

$$\mathbf{K}_{11} = \frac{\sqrt{2}}{L^3} \begin{bmatrix} C_1 & 0 & 0 \\ 0 & C_1 & 0 \\ 0 & 0 & 2C_2 L^2 \end{bmatrix} \quad (20)$$

$$\mathbf{K}_{12} = \sqrt{2}\alpha_2^2 L^5 \begin{bmatrix} -EAL^4 + 6C_3 R^2 & 2\sqrt{2}C_4 LR\alpha_1 & C_5 L^2 R\alpha_2 \\ 2\sqrt{2}C_4 LR\alpha_1 & -4C_6 R^2 - 6EJL^2 & -\sqrt{2}C_7 L^3 \alpha_2 \\ -C_5 L^2 R\alpha_2 & \sqrt{2}C_7 L^3 \alpha_2 & C_8 L^4 \alpha_2^2 \end{bmatrix} \quad (21)$$

$$\mathbf{K}_{13} = \sqrt{2}\alpha_2^2 L^5 \begin{bmatrix} -4C_6 R^2 - 6EJL^2 & -2\sqrt{2}C_4 LR\alpha_1 & \sqrt{2}C_7 L^3 \alpha_2 \\ -2\sqrt{2}C_4 LR\alpha_1 & -EAL^4 + 6C_3 R^2 & C_5 L^2 R\alpha_2 \\ -\sqrt{2}C_7 L^3 \alpha_2 & -C_5 L^2 R\alpha_2 & C_8 L^4 \alpha_2^2 \end{bmatrix} \quad (22)$$

$$\mathbf{K}_{14} = \sqrt{2}\alpha_2^2 L^5 \begin{bmatrix} -EAL^4 + 6C_3 R^2 & 2\sqrt{2}C_4 LR\alpha_1 & -C_5 L^2 R\alpha_2 \\ 2\sqrt{2}C_4 LR\alpha_1 & -4C_6 R^2 - 6EJL^2 & \sqrt{2}C_7 L^3 \alpha_2 \\ C_5 L^2 R\alpha_2 & -\sqrt{2}C_7 L^3 \alpha_2 & C_8 L^4 \alpha_2^2 \end{bmatrix} \quad (23)$$

$$\mathbf{K}_{15} = \sqrt{2}\alpha_2^{\,2}L^5 \begin{bmatrix} -4C_6R^2 - 6EJL^2 & -2\sqrt{2}C_4LR\alpha_1 & -\sqrt{2}C_7L^3\alpha_2 \\ -2\sqrt{2}C_4LR\alpha_1 & -EAL^4 + 6C_3R^2 & -C_5L^2R\alpha_2 \\ \sqrt{2}C_7L^3\alpha_2 & C_5L^2R\alpha_2 & C_8L^4\alpha_2^{\,2} \end{bmatrix} \quad (24)$$

$$\mathbf{K}_{22} = \sqrt{2}\alpha_2^{\,2}L^5 \begin{bmatrix} EAL^4 - 6C_3R^2 & -2\sqrt{2}C_4LR\alpha_1 & -C_5L^2R\alpha_2 \\ -2\sqrt{2}C_4LR\alpha_1 & 4C_6R^2 + 6EJL^2 & \sqrt{2}C_7L^3\alpha_2 \\ -C_5L^2R\alpha_2 & \sqrt{2}C_7L^3\alpha_2 & C_2L^4\alpha_2^{\,2} \end{bmatrix} \quad (25)$$

$$\mathbf{K}_{33} = \sqrt{2}\alpha_2^{\,2}L^5 \begin{bmatrix} 4C_6R^2 + 6EJL^2 & 2\sqrt{2}C_4LR\alpha_1 & -\sqrt{2}C_7L^3\alpha_2 \\ 2\sqrt{2}C_4LR\alpha_1 & EAL^4 - 6C_3R^2 & -C_5L^2R\alpha_2 \\ -\sqrt{2}C_7L^3\alpha_2 & -C_5L^2R\alpha_2 & C_2L^4\alpha_2^{\,2} \end{bmatrix} \quad (26)$$

$$\mathbf{K}_{44} = \sqrt{2}\alpha_2^{\,2}L^5 \begin{bmatrix} EAL^4 - 6C_3R^2 & -2\sqrt{2}C_4LR\alpha_1 & C_5L^2R\alpha_2 \\ -2\sqrt{2}C_4LR\alpha_1 & 4C_6R^2 + 6EJL^2 & -\sqrt{2}C_7L^3\alpha_2 \\ C_5L^2R\alpha_2 & -\sqrt{2}C_7L^3\alpha_2 & C_2L^4\alpha_2^{\,2} \end{bmatrix} \quad (27)$$

$$\mathbf{K}_{55} = \sqrt{2}\alpha_2^{\,2}L^5 \begin{bmatrix} 4C_6R^2 + 6EJL^2 & 2\sqrt{2}C_4LR\alpha_1 & \sqrt{2}C_7L^3\alpha_2 \\ 2\sqrt{2}C_4LR\alpha_1 & EAL^4 - 6C_3R^2 & C_5L^2R\alpha_2 \\ \sqrt{2}C_7L^3\alpha_2 & C_5L^2R\alpha_2 & C_2L^4\alpha_2^{\,2} \end{bmatrix} \quad (28)$$

with the auxiliary coefficients $\alpha_1 = \sqrt{1 - \dfrac{6R^2}{L^2}}$ and $\alpha_2 = \sqrt{1 + \dfrac{2R^2}{L^2}}$ and the auxiliary parameters

$$\begin{aligned} C_1 &= EAL^2 + 6EJ \\ C_2 &= EAR^2 + 4EJ \\ C_3 &= EAL^2 - 8EJ \\ C_4 &= EAL^2 - 6EJ \\ C_5 &= EAL^2\alpha_1 - 12EJ \\ C_6 &= 2EAL^2 - 9EJ \\ C_7 &= 2EAR^2 + 3EJ\alpha_1 \\ C_8 &= EAR^2 + 2EJ \end{aligned} \quad (29)$$